\documentclass[pre,twocolumn]{revtex4-1} 
\usepackage{amsmath}
\usepackage{amsfonts}
\usepackage{graphicx}
\usepackage{comment}
\usepackage{bm}
\usepackage{braket}
\DeclareMathOperator{\Tr}{Tr}

\begin{document}

\title{Local structure characterization in particle systems}
\author{R. S. Skye}
\email{rs136@wellesley.edu}
\author{E. G. Teich}
\email{et106@wellesley.edu}
\affiliation{Department of Physics and Astronomy, Wellesley College, Wellesley, Massachusetts 02481}

\date{\today}

\begin{abstract}
Many tools and techniques measure local structure in materials in contexts ranging from biology to geology.
We provide a survey of those tools and metrics  that are especially useful for analyzing particulate soft matter.
The metrics we discuss can all be computed from the positions of particles, and are thus most useful when there is access to this information, either from simulation or experimental imaging.
For each metric, we provide derivations, intuition regarding its implications, example uses, and references to software packages that compute the metric.
Our survey encompasses characterization techniques ranging from the simplest to the most complex, and will be useful for students getting started in the structural characterization of particle systems.
\end{abstract}

\maketitle

\section{Introduction}
Our natural and human-made world is formed from materials with a staggering range of structural organization.
Often, these materials can be decomposed into ``particles" of some type, including angstrom-scale atoms,\cite{Rohrer2001} nano-scale molecules or proteins,\cite{Kitaigorodsky1973} micro-scale colloids or cells,\cite{Li2011} millimeter-scale seeds,\cite{Armon2014} and macro-scale grains,\cite{Behringer2019} to name a few examples.
The arrangement, or structure, of these particles can be complex, messy, or noisy. 
Nevertheless, an accurate understanding of any material's structure is crucial for understanding its properties and how it can be engineered for new technologies.

In this paper, we introduce a suite of computational tools for the characterization of \emph{local particle environments} in generic particulate systems.
Characterizing local neighborhoods around particles is a powerful and general means of probing the structure of any material, whether it is a periodically ordered crystal,\cite{Jones2010, Blatova2023} completely disordered material,\cite{Bernal1959, Royall2015, Yang2021c} or anywhere in between.\cite{Takakura2007, Zhou2024, Keen2015}
Additionally, monitoring local neighborhoods and how they dynamically change enables a clearer understanding of crystallization,\cite{Je2021, Martirossyan2024} the glass transition,\cite{Leocmach2012, Royall2015} transitions between crystal structures,\cite{Marcus1996, Du2017} and the evolution of structure under external forcing.\cite{Schlegel2016,Richard2020a}
To illustrate the utility of these metrics, we will present several examples in soft matter (and particularly, simulated soft matter systems) because that is our area of expertise.
However, the techniques described can be used in many contexts, ranging from the structure of biological tissues\cite{Tang2024} to the distribution of galaxies in the universe.\cite{Jones2005}
Almost all of the tools we will describe are implemented in multiple software packages dedicated to the analysis and visualization of particle systems generated by simulations.
We will highlight the utility of four packages in particular: \texttt{freud},\cite{Ramasubramani2020} \texttt{OVITO},\cite{Stukowski2010} \texttt{pyscal},\cite{Menon2019} and \texttt{mdapy}. \cite{Wu2023}
Each has a Python API and is actively maintained, and may therefore be especially helpful for students.
Table~\ref{table:methods} lists the methods used by each program to calculate every metric we introduce.

\begin{table*}[t]
\centering
\begin{footnotesize}
\begin{tabular}{lllll}
\hline
\textbf{Metric} & \textbf{freud} & \textbf{OVITO} & \textbf{pyscal} & \textbf{mdapy} \\
\hline
\hline
$CN$ & \texttt{locality.LinkCell} & \texttt{Coordination analysis} & \texttt{find\_neighbors} & \texttt{neighbor} \\
$g(r)$ & \texttt{density.RDF} & \texttt{Coordination analysis} & \texttt{calculate\_rdf} & \texttt{pair\_distribution} \\
BOOD & \texttt{environment.BondOrder} & & & \\
$\psi_\ell$ & \texttt{order.Hexatic} & &  & \\
$Q_\ell$ & \texttt{order.Steinhardt} & & \texttt{calculate\_q} & \texttt{steinhardt\_bond\_orientation} \\
env & \texttt{environment.EnvironmentCluster} &  &  & \\
CNA & & \texttt{Common neighbor analysis} & \texttt{calculate\_cna} & \texttt{common\_neighbor\_analysis} \\
PTM & & \texttt{Polyhedral template matching} & & \texttt{polyhedral\_template\_matching} \\
\hline
\end{tabular}
\end{footnotesize}
\caption{These software packages implement the structural characterization techniques we discuss.
Each metric is listed in the left-most column in its abbreviated form, in the order it is introduced.
Note that ``BOOD" abbreviates bond orientational order diagram, ``env" abbreviates environment matching, ``CNA" abbreviates common neighbor analysis, and ``PTM" abbreviates polyhedral template matching.
Methods to calculate each metric are listed under their respective software package.
All methods listed in the \texttt{OVITO} column are pipeline modifiers.
Blank entries indicate that the software package does not support the calculation of the indicated metric.}
\label{table:methods}
\end{table*}

\section{What is a local environment?} \label{section:definition}
\begin{figure}[ht!]
    \centering
    \includegraphics[width=0.5\textwidth]{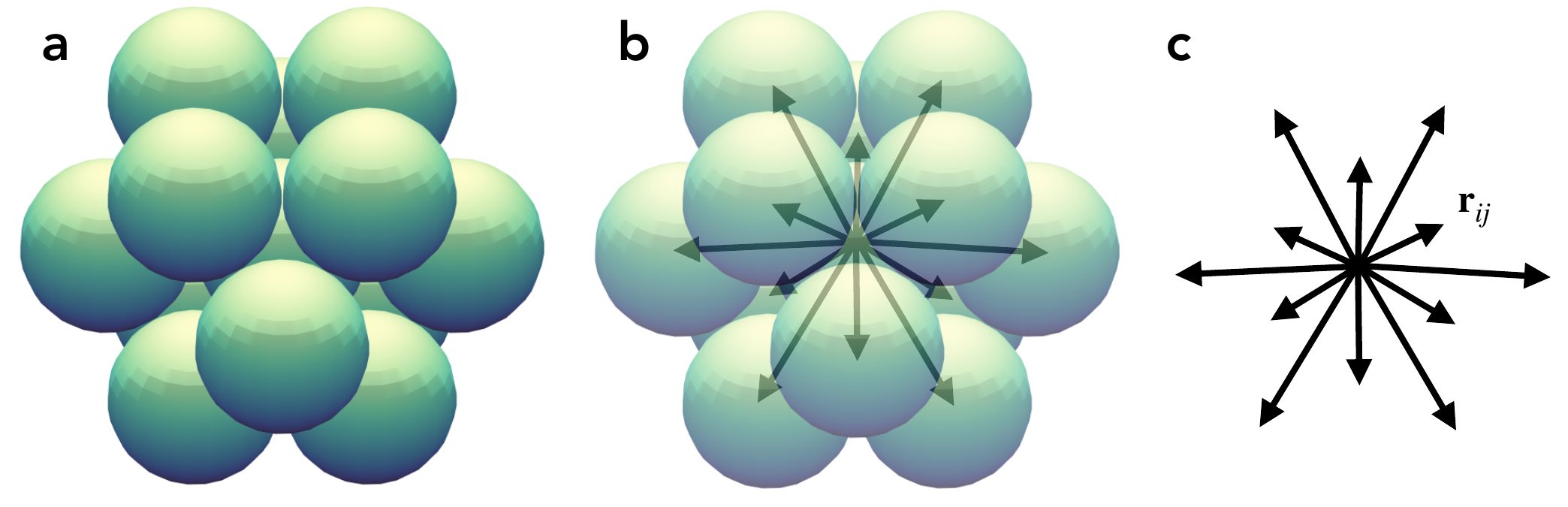}
    \caption{
    The extraction of an environment vector set from a local neighborhood.
    (a) For any set of particle neighbors near a central particle, (b) a set of vectors can be defined pointing from the center of the central particle $i$ to the centers of its neighbors, indexed by $j$.
    (c) This vector set, denoted as $\{ \mathbf{r}_{ij} \}$, can be viewed as the abstract representation of any local environment.
    }
    \label{fig:env}
\end{figure}

We define the local environment surrounding particle $i$ as the set of all vectors pointing from the center of $i$ to the centers of its nearest neighbors.
We call this set of vectors $\{ \mathbf{r}_{ij}\}$, where $j$ indexes over all nearest neighbors and $\mathbf{r}_{ij}$ points from the center of particle $i$ to the center of particle $j$.
(This vector is sometimes referred to as a ``bond," regardless of the type of interaction between the particles.)
For any particle $i$, which nearby particles do we call its ``nearest neighbors"?
There is no universally correct answer: nearest neighbors can be defined in different ways depending on the needs of the analysis.
Two common definitions are the set of $N$ closest particles to particle $i$, and all particles within some distance $r_{\max}$ of particle $i$. 
Either $N$ or $r_{\max}$ are chosen by the user.
There exist more complex, and often more robust, methods for identifying nearest neighbors; for example, the Voronoi construction finds the polyhedral region of space that is closer to particle $i$ than to any other particle and identifies the particles $j$ that are equidistant with $i$ to these facets as the nearest neighbors. \cite{Lazar2022}
Figure~\ref{fig:env} shows the extraction of the environment vector set $\{ \mathbf{r}_{ij}\}$ from an example local particle neighborhood.

Each vector $\mathbf{r}_{ij}$ within a local environment has an associated magnitude and direction, making each environment a $2N$ or $3N$-dimensional signal (where $N$ is the number of neighbors in each environment) for systems in 2 or 3 dimensions, respectively.
Each particle in the system has its own unique environment, and thus analyzing local structure in a system becomes a problem of collapsing many high-dimensional signals into interpretable and useful metrics.
These metrics must also be robust against the inevitable  distortion that results from thermal noise and particle tracking errors in experiments.
We will see that the techniques we explore utilize varying amounts of information about each local environment to characterize structure.
The following sections are roughly organized so that each newly-introduced technique uses more structural information than the one preceding it.

\section{Coordination number} \label{section:CN}

A simple measure of local structure is the coordination number $CN$, the number of members in the set $\{ \mathbf{r}_{ij}\}$. 
This number can be reported as an average for the entire system or a histogram. 
However, within this simple definition lies complexity.

\begin{figure}[h!]
    \centering
    \includegraphics[width=0.5\textwidth]{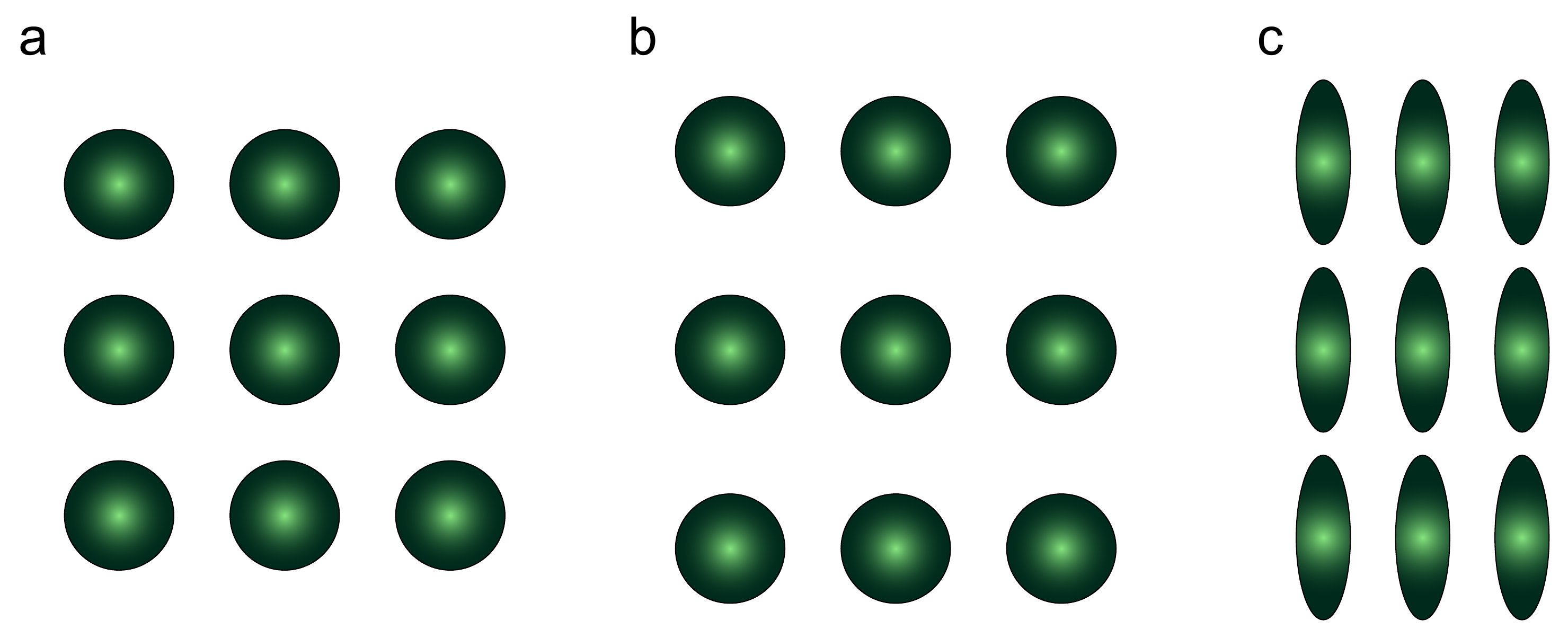} 
    \caption{
    Examples of coordination environments which may require different definitions of the local neighborhood. 
    (a) A square lattice has a single simple nearest-neighbor distance,
    (b) a rectangular lattice may be defined as having two nearest neighbor distances, and 
    (c) a lattice of ellipsoids may have a complex definition of the distance to a neighbor.
    }
    \label{fig:neighborhood}
\end{figure}

A measure of the nearest neighbors of a particle necessitates a definition of a nearest neighbor. 
We previously discussed several definitions of a local environment. 
Restricting the system to $N$ nearest neighbors does not provide any information, and thus we must choose  $r_{\max}$ to define the local environment. 
In simple cases such as a square lattice, there is a simple cut-off between the nearest neighbors and second-nearest neighbors [see Fig.~\ref{fig:neighborhood}(a)]. 
If the nearest neighbors are approximately at distance $a$, then the second-nearest neighbors are at distance $\sqrt{2}a$. 
Thus, setting $r_{\max}$ equal to any distance between $a$ and $\sqrt{2}a$ will result in $CN=4$ for each particle in the lattice.
In other systems, however, the most useful $r_{\max}$ is not so obvious. 
In a rectangular lattice, a particle may have two neighbors at distance $a$ and two at $1.1a$ [see Fig.~\ref{fig:neighborhood}(b)]. 
Whether or not both sets are counted in the coordination number may depend on the application. 
Anisotropic particles also complicate the definition of the coordination number. 
For example, a system composed of elongated ellipsoids contains two types of neighbors: those along the short axis, and those along the long axis [see Fig.~\ref{fig:neighborhood}(c)]. 
These two groups have different characteristics and may sit at very different distances from the particle origin, necessitating a more thoughtful method of calculating nearest-neighbors than a simple $r_{\max}$.

One method of counting nearest neighbors that is useful for disordered materials involves constructing the Voronoi tessellation of the system. 
A Voronoi polyhedron around a reference particle defines the region of space that is closer to the particle than to any other particle.\cite{Lazar2022}
The number of facets on the polyhedron are then taken as the coordination number. 
A thorough discussion of Voronoi tessellation is given in Ref.~\onlinecite{Lazar2022}.
The software packages listed in Table~\ref{table:methods}  can be used to find nearest neighbors using this technique.

In simple structures, all particles have the same coordination number (apart from defects). 
However, for more complex structures this is not the case. 
Consider, for instance, SiO$_{2}$ quartz. 
Silicon atoms have a preferred 4-coordination, and oxygen atoms a preferred 2-coordination.\cite{dAmour1979}
To say that the system has coordination number $\frac{1}{3} \times 4+\frac{2}{3}\times 2=2.67$ is not reflective of either local environment. 
In these situations, a histogram of coordination numbers or site-specific coordination numbers are more informative.

Another difficulty when calculating coordination numbers is the existence of a surface. 
In  simple cases, a surface particle has approximately half the number of neighbors as a bulk particle, and thus strongly skews the observed coordination. 
A simple method of discounting surface particles is to group those with a particularly small $CN$ and exclude them from the analysis.

We can gain intuition on whether an environment is  high- or low-coordinated by considering simple sphere systems. 
For two-dimensional (2D) systems, close-packing of spheres (disks) results in a hexagonal structure with $CN=6$.\cite{Nelson1989}
Thus, dense 2D systems of particles with isotropic interactions often have $CN\sim 6$.
For three-dimensional (3D) systems, close-packing of spheres results in an icosahedral local environment with $CN=12$,\cite{Nelson1989} 
and thus dense 3D systems of isotropic particles often have $CN\sim 12$, with some local environments reaching $CN\sim14$ or $15$.\cite{Dshemuchadse2022}
Systems of particles with anisotropic interactions can have lower $CN$, with $CN=4$ the lowest typical possible value in connected 3D crystals. \cite{Dshemuchadse2022}

\section{Radial distribution function} \label{section:rdf}

The radial distribution function, $g(r)$, measures fluctuations in the density moving outward from an average reference particle.\cite{Chandler1987} 
It is the average number density as a function of the distance $r$ from any particle center, relative to the mean overall density of the system. 
Short-range, medium-range, and long-range order can be extracted from $g(r)$,\cite{Branka2011} unlike the coordination number, which can typically only be used to understand short-range order. 
Notably, no angular information is retained:  $g(r)$ is an average over all directions. 

\begin{figure}[h!]
    \centering
    \includegraphics[width=0.4\textwidth]{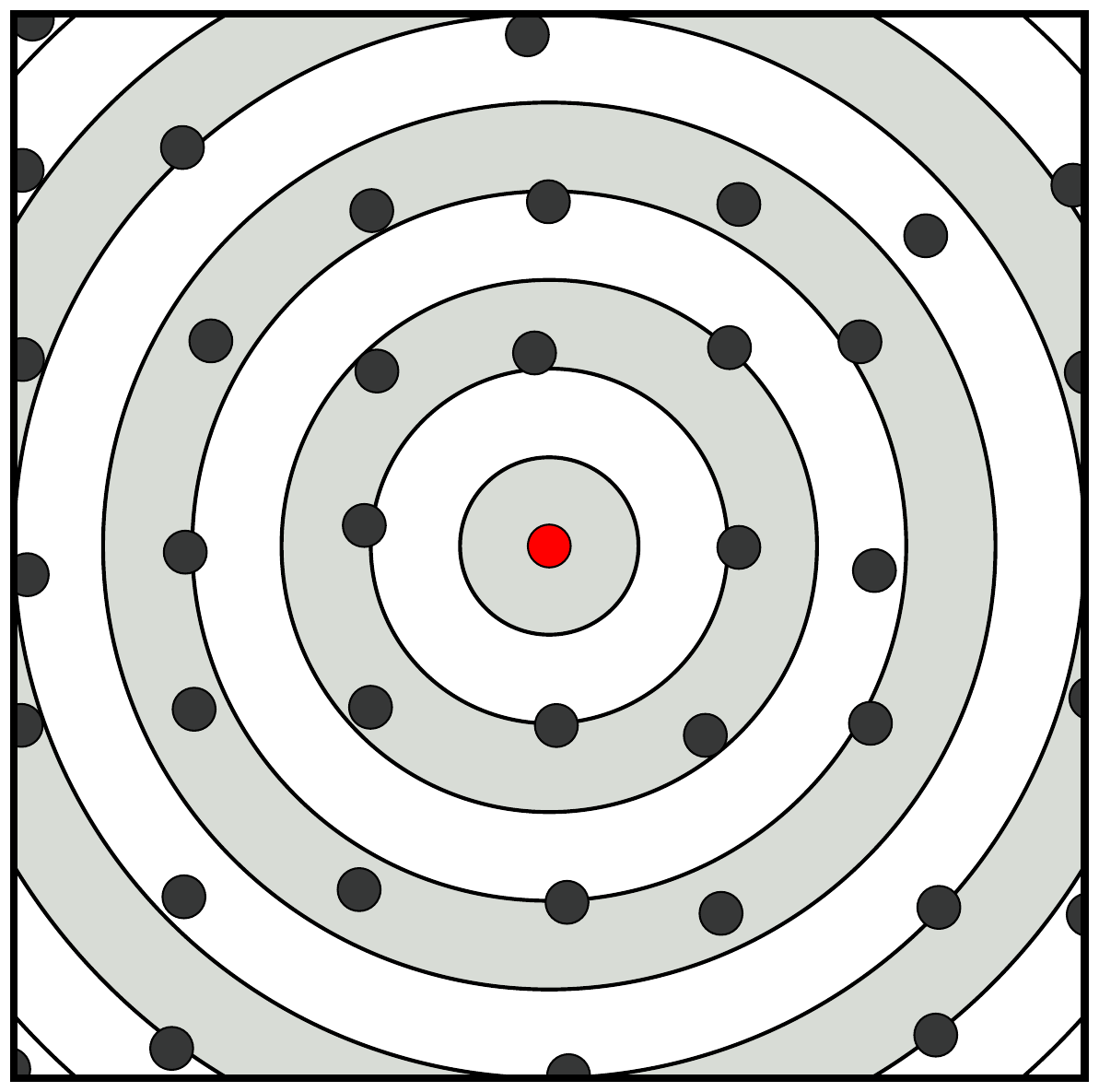} 
    \caption{
    A schematic representation for calculating $g(r)$ for a a square lattice with noise. 
    The red particle indicates the central reference particle. 
    $g(r)$ is constructed by calculating the number density of particles within each shell moving outward, and comparing to the overall density. 
    This process is repeated taking each particle as the central reference to calculate an average.
    }
    \label{fig:rdfshells}
\end{figure}

To construct $g(r)$,\cite{Frenkel2002} we first measure all distances $|\mathbf{r}_{ij}|$ between particle $i$ and all other particles within a large cut-off distance, typically much larger than just the nearest-neighbor distance.
These distances are collected for all particles $i$, and a histogram of distances is computed for thin circular (2D) or spherical (3D) ``shells'' of thickness $dr$ (Fig.~\ref{fig:rdfshells}). 
Choosing $dr$ is important, as a thick shell will average out potentially useful information but a thin shell will be overly sensitive to system noise.
Once the histogram has been constructed, common interparticle distances appear as peaks.   

Considering only the number of particles in a shell yields the \textit{pair distribution function} which increases as $r$ increases.
To avoid an infinite increase, we instead calculate the number density, or the number of particles per unit volume of a thin shell. 
This number is then normalized by the mean density of the entire system to compute $g(r)$, so that $g(r) = 1$ if the number density at $r$ is equal to the average number density of the system. 
By this construction, for disordered or noisy systems $g(r)$ will approach unity at long distances.

If $dn(r)$ is the number of neighbor particles in a small range $dr$ around $r$, then $g(r)$ is defined as
\begin{align}
    g(r) = \frac{dn(r)}{dV(r)~\rho}~,
\end{align}
where $\rho$ is the mean number density and $dV(r)$ is the volume of a thin spherical shell of radius $r$ and thickness $dr$. 
Shells are typically considered thin enough to use the thin-wall approximation given in 3D by 
\begin{align}
    g(r) = \frac{dn(r)}{4 \pi r^2 dr~\rho}.
\end{align}

This definition is most useful in simulations, or in systems which can be accurately visualized under a microscope, because it requires knowing the location of each particle precisely. 
An alternative definition, useful for comparison to experimental values, uses the structure factor (which is related to the Fourier transform of $g(r)$ and can be determined from scattering experiments) to calculate $g(r)$. Interested readers can refer to Ref.~\onlinecite{Yarnell1973} for more detail.

Several useful measures of structure can be extracted from $g(r)$. 
The coordination number (Sec.~\ref{section:CN}) can be calculated by integrating $4\pi r^2 \rho g(r)$ over the first peak of $g(r)$,\cite{Cristiglio2009} and the average distance to these nearest neighbors can be estimated as the location of this first peak.
Often the distance to the first minimum after the first peak is chosen as the definition of $r_{\max}$, the cut-off distance that defines each particle's local environment.
Additional neighbor-shells after the first can be defined to examine second-nearest-neighbors and beyond.\cite{Cristiglio2009}

\begin{figure}[h!]
    \centering
    \includegraphics[width=0.4\textwidth]{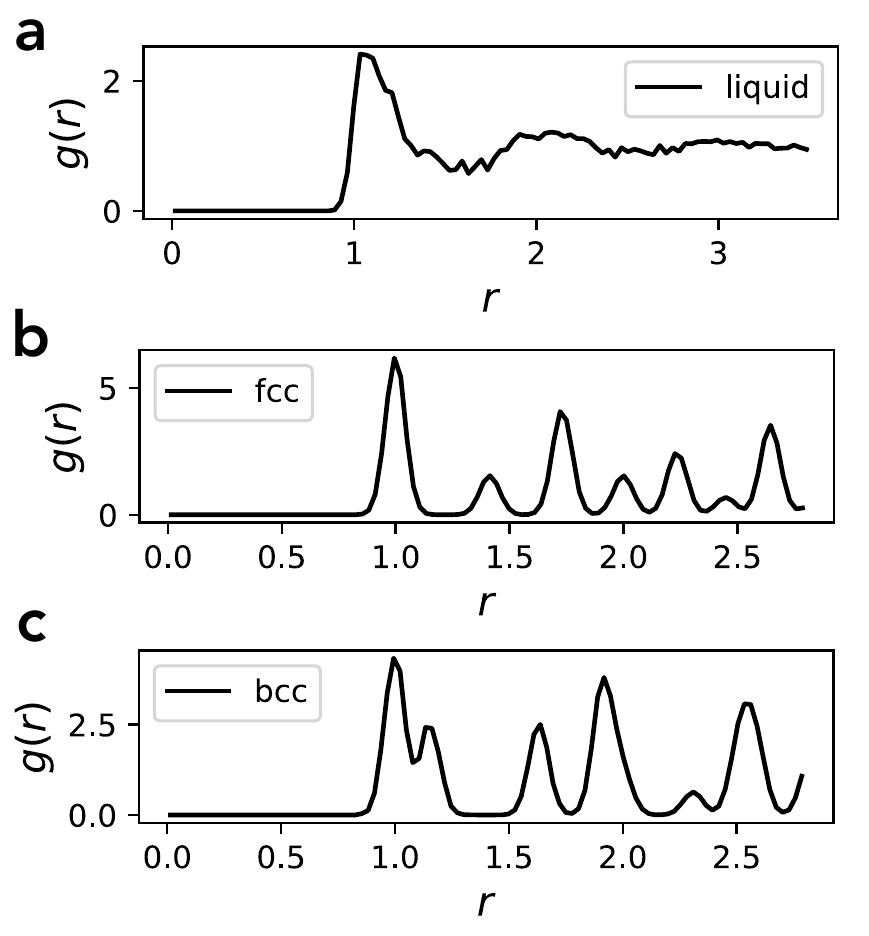} 
    \caption{
    Plots of $g(r)$ for (a) liquid, (b) fcc crystal, and (c) bcc crystal structures.
    The liquid is generated from 500 particles interacting via the Lennard-Jones potential discussed in Sec.~\ref{section:detectingcrystals}, with parameters $\varepsilon = 1$ and $\sigma = 1$.
    The system was simulated via molecular dynamics\cite{Anderson2020} in the NVT ensemble at $k_B T = 1.2$ and number density $\rho = 0.8$; $g(r)$ was calculated from a single system snapshot after equilibration.
    Each crystal structure consists of 10 replicated unit cells using the \texttt{UnitCell} class within \texttt{freud}, with Gaussian noise of standard deviation 0.03 added to each particle position.
    Unit cells were scaled so that the nearest-neighbor distance is unity in both structures.
    }
    \label{fig:rdf2}
\end{figure}

The shape of $g(r)$ is often used to identify the phase of a system.\cite{Huong2014} 
For an ideal gas, $g(r) = 1$. 
For a real gas, there is a minimum distance---usually correlated with the particle diameter---below which $g(r) \sim 0$. 
Above this distance $g(r)$ increases rapidly, beyond $g(r)=1$, then decays to unity exponentially as $r$ increases. 
For a liquid or disordered solid, there is a minimum distance before the first peak, followed by long-range decaying oscillations to $g(r)=1$. 
These fluctuations can be used to estimate the correlation length of local order\cite{Branka2011} by calculating the distance over which the peak maxima decay to within some fraction of the global maximum. 
For an ordered solid, $g(r)$ is a ``fingerprint'' of peaks that can be used to identify the structure. 
In a perfectly ordered solid, these peaks would be delta functions and would never decay to precisely $g(r) = 1$. 
In a real solid, the peaks exhibit a finite width due to thermal noise, and $g(r)$ decays to $1$ as defects and grain boundaries compound at long distances. 
Figure~\ref{fig:rdf2} shows  $g(r)$ for a liquid, the face-centered cubic (fcc) crystal structure and the body-centered cubic (bcc) crystal structure.
The $g(r)$ distribution for the liquid shows a characteristic prominent first peak, representing the nearest neighbor shell, followed by decaying oscillations at larger values of $r$ [Fig.~\ref{fig:rdf2}(a)].
In contrast, $g(r)$ distributions for the fcc [Fig.~\ref{fig:rdf2}(b)] and bcc [Fig.~\ref{fig:rdf2}(c)] crystal structures show narrower, well-separated peaks at distinct values of $r$, representing the interparticle distances in each crystal.
The differences between these $g(r)$ signatures reflect structural differences between the crystals.
The width of each peak is nonzero due to the presence of noise.

In multi-component systems, $g(r)$  can be computed for each component separately to reveal information about the relation between the components. 
For instance, in a binary system of components A and B, we could calculate $g(r)$ for only A-type particles, only B-type, or only the density of B particles measured outward from A particles. 
These distributions provide information about the relative order of these components: is the nearest-neighbor-shell composed of like particles or unlike particles? 
Is the A component ordered but the B component disordered? 
Are the correlation lengths of the two components different?

\section{Bond orientational order} \label{section:BOOD}

\begin{figure}[h]
    \centering
    \includegraphics[width=0.5\textwidth]{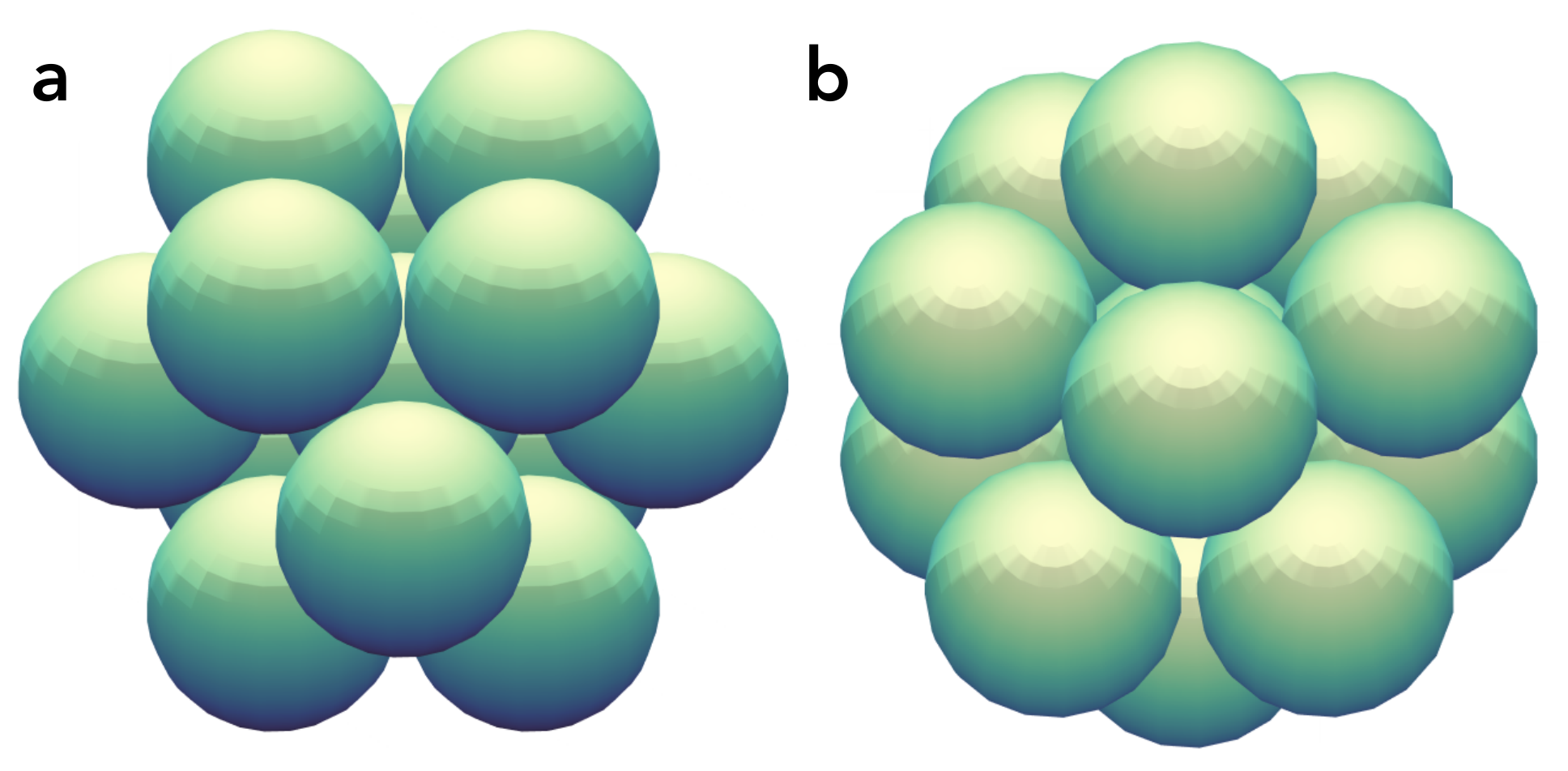} 
    \caption{
    Local environments may differ primarily in their rotational symmetry.
    (a) The fcc local environment and (b) the icosahedral environment both consist of 12 particles packed tightly around a central particle, but have significantly different rotational symmetry.
    }
    \label{fig:envII}
\end{figure}

The radial distribution function considers only the distance between particles with no consideration of the \textit{orientation} of the bonds. 
Other metrics analyze these bond orientations to provide important information about the rotational symmetry of local environments. \cite{Roth1995,Roth2000,Damasceno2012, Engel2015}
Rotational symmetry is very useful for distinguishing between two local environments that are otherwise similar according to other measurements.
For example, the local neighborhood associated with the fcc structure consists of 12 particles that are closely packed around a single center particle in a cuboctahedral arrangement [Fig.~\ref{fig:envII}(a)].
In contrast, the local neighborhood associated with close-packed disorder in 3D consists of 12 particles that are closely packed around a single center particle in an icosahedral arrangement [Fig.~\ref{fig:envII}(b)].
Both local environments have the same coordination number and similar $g(r)$ distributions.
They are distinct arrangements, however, with the fcc environment having 2-fold, 3-fold, and 4-fold rotational symmetries, and the icosahedral environment having 2-fold, 3-fold, and 5-fold rotational symmetries.
Analyzing these environments' bond orientational order is useful for distinguishing between them.

To visualize bond orientational order, we can construct a bond orientational order diagram as follows:
\begin{enumerate}
    \item For each particle $i$, compute the angles of all bonds in its local environment $\{\mathbf{r_{ij}}\}$. 
    In 2D, each environment vector is specified by the angle $\theta_{ij}$ that it makes with respect to some reference line.
    In 3D, each environment vector can be specified by two angles, $\theta_{ij}$ (polar) and $\phi_{ij}$ (azimuthal).
    \item Construct a histogram of all bond angles in the system. 
    \item Visualize the bond orientational order diagram directly as a histogram (over the angle $\theta$ in 2D or over the angles $\phi$ and $\theta$ in 3D), or visualize it as a heat map over the circumference of the unit circle (2D) or the surface of the unit sphere (3D).
\end{enumerate}

\begin{figure}[h!]
    \centering
    \includegraphics[width=0.5\textwidth]{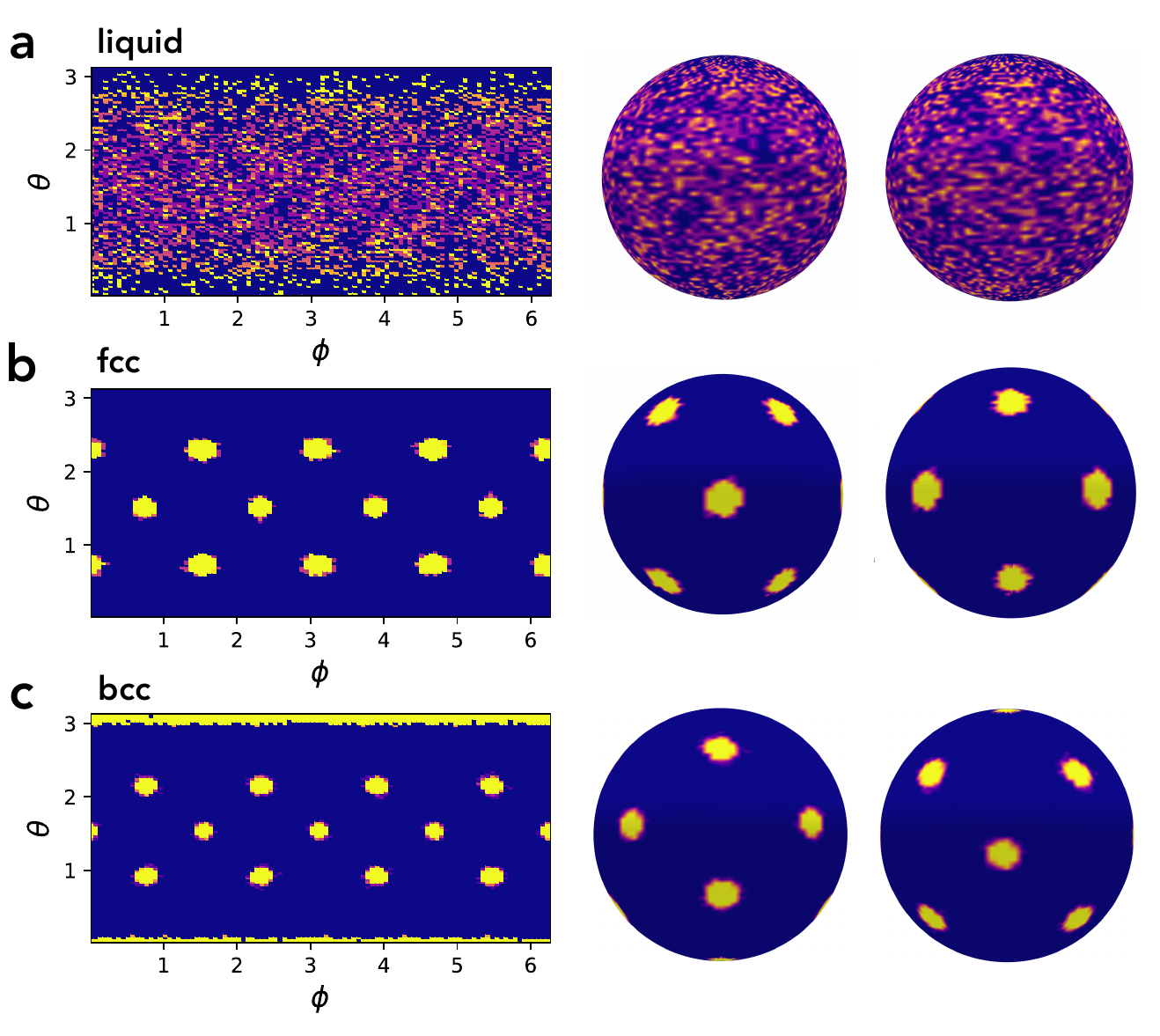} 
    \caption{
    Example bond orientational order diagrams for (a) liquid, (b) fcc crystal, and (c) bcc crystal structures.
    For each structure, the bond orientational order diagram is shown as a 2D histogram over the angles $\phi$ and $\theta$ in the left panel, and as a heat map over the surface of the unit sphere in the middle and right panels. 
    The middle and right panels display two rotated views of the spherical heat map, to show two axes of rotational symmetry for each structure. 
    The bond orientational order diagram values are clipped to the 95th percentile for the liquid and fcc structures, and to the 90th percentile for the bcc structure, to aid visualization.
    Neighbors are defined as those within the distance $r_{\max} = 1.5$ for the liquid, $r_{\max} = 1.2$ for  the fcc structure, and $r_{\max} = 1.4$ for the bcc structure.
    Each distance roughly corresponds to the distance of the minimum after the first peak of the corresponding $g(r)$.
    }
    \label{fig:bod}
\end{figure}

If angles are measured with respect to global axes, information is retained about orientational order in the system as a whole. 
In a system such as a liquid or gas, the histogram will be isotropic and featureless. 
In an ordered system with only one grain, the histogram will exhibit strong peaks that indicate the symmetry of all particles' local environments, superimposed.
Figure~\ref{fig:bod} shows bond orientational order diagrams for the systems used to calculate the $g(r)$ distributions in Fig.~\ref{fig:rdf2}.
Each  diagram is shown in a 2D histogram form and as a pattern on the surface of the unit sphere.
Note the isotropic, featureless nature of the bond orientational order diagram corresponding to the liquid structure [Fig.~\ref{fig:bod}(a)].
In contrast, the bond orientational order diagrams for the fcc [Fig.~\ref{fig:bod}(b)] and bcc [Fig.~\ref{fig:bod}(c)] crystal structures display clear and distinct peaks.
The peaks show the cuboctahedral (for fcc) and rhombic dodecahedral (for bcc) symmetry of each local environment.

Several aspects of the system can render the corresponding bond orientational order diagram more complicated to interpret. 
If the system contains multiple grains or significant defects, they will appear as peaks shifted by their relative orientation. 
In a large system with many randomly oriented grains, the bond orientational order diagram will appear isotropic and indistinguishable from that for a disordered system. 
If there are only a small number of grains, there may be multiple distinct sets of peaks. 
In this case, it is common to augment the analysis with other order parameters, because there is no general method for distinguishing between the case of multiple grains and the case of a complex local environment. 
For instance, stacking faults in the fcc crystal structure (in which an entire layer of the structure does not conform to the overall crystalline pattern) are similar in local environment to the hexagonally close-packed (hcp) crystal structure, and thus spatially resolved methods must be used to determine if the signature is produced by a defect or by a separate grain.\cite{Rohrer2001}
(For an example of a bond orientational order diagram produced by fcc stacking faults, see Fig.~\ref{fig:LJ1}(d), right).
If the system is textured---\textit{i.e.}, it contains many grains that adopt a preferred subset of all possible orientations---or exhibits liquid crystalline order, then peaks will instead be streaks or rings corresponding to the allowed orientations.\cite{Damasceno2012}

If simultaneously observing all bond orientations of all environments is not useful, then there exist two possibilities. 
First, we can select a subset of the system which contains only a single grain for analysis. 
This selection can be done manually, but is difficult and time consuming for large data processing. 
Second, each $(\theta_{ij},\phi_{ij})$ can instead be measured relative to a \textit{local} reference frame. 
This reference frame can be with respect to the orientation of an anisotropic particle, which can function to align grains in dense systems of strongly oriented particles.\cite{Damasceno2012, Ramasubramani2020}
However, in systems of isotropic particles the reference particle orientation is not meaningful, and not all assemblies of anisotropic particles are ordered with respect to the individual particle.
Some structural analysis methods, such as environment matching (Sec.~\ref{section:env}) or polyhedral template matching\cite{Larsen2016} (Sec.~\ref{section:additional}), capture the local environment orientation as part of their algorithm. 
This orientation could be fed back as an input to a bond orientational order diagram calculation to orient a local structure, provided it is well-ordered.

The bond orientational order diagram is useful for differentiating structures that are similar with respect to $g(r)$ by distinguishing them according to their local symmetry. 
It is also invaluable for identifying quasicrystals\cite{Roth1995,Roth2000,Engel2015,Je2021} because ``forbidden'' rotational symmetry axes (\emph{e.g.}, 5-, 10-, and 12-fold) will be apparent. 
Nonetheless, the technique must be used with caution, because defects and in particular stacking faults can also be observed as apparently ``forbidden'' symmetries.\cite{Mackay1962}
The software package \texttt{INJAVIS}\cite{injavis} is especially useful for visualizing bond orientational order diagrams of crystal structures: In the \texttt{INJAVIS} environment, structures and bond orientational order diagrams can be rotated simultaneously, and the bond orientational order diagram can be calculated instantaneously from a set of bonds by highlighting a range of bond distances within an auto-computed $g(r)$ distribution.  

\section{Detecting crystallization}\label{section:detectingcrystals}

\begin{figure}[h!]
    \centering
    \includegraphics[width=0.5\textwidth]{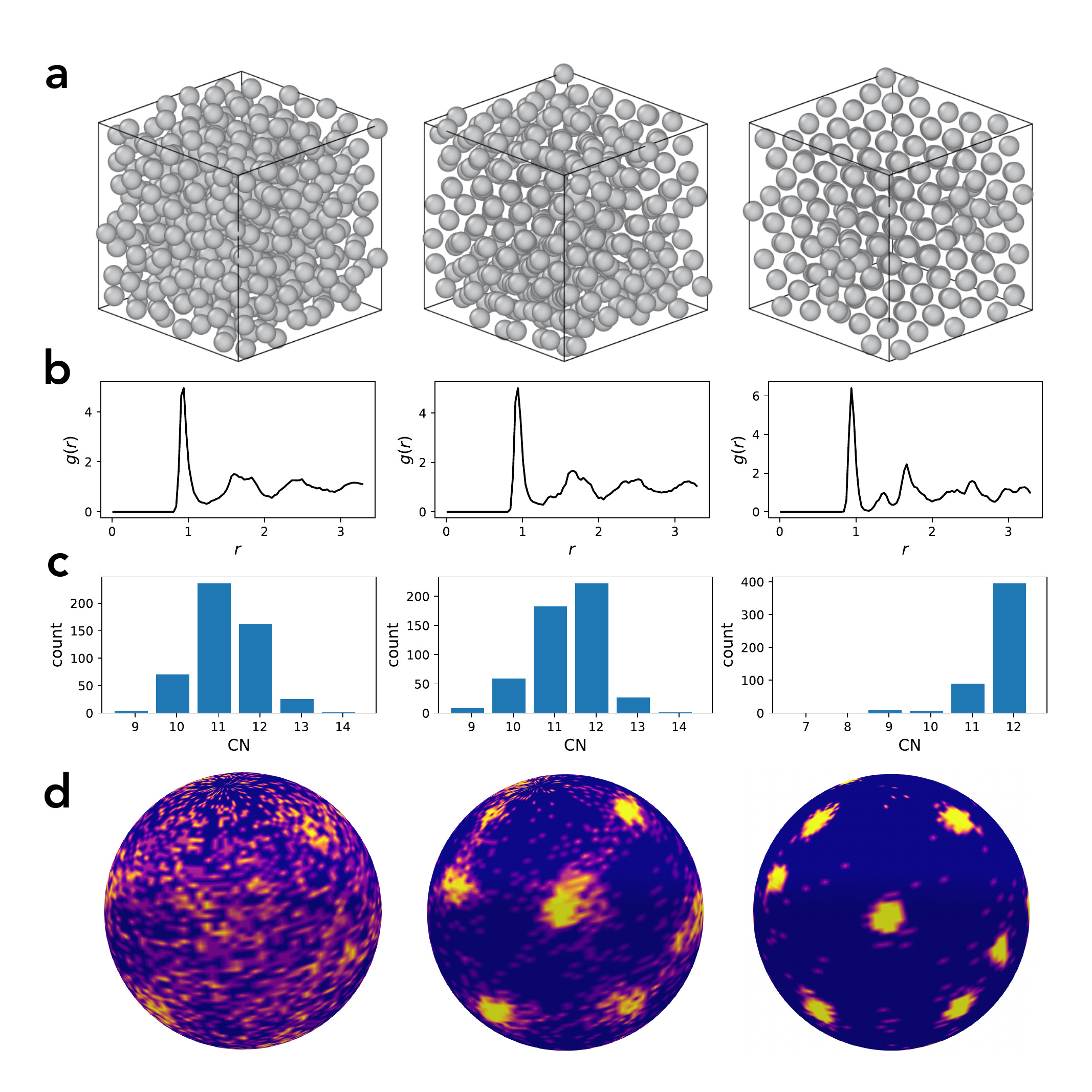} 
    \caption{
    Crystallization in the Lennard-Jones system is indicated by changes in $g(r)$, the coordination number distribution, and the bond orientational order diagram.
    The system consists of 500 particles with $U(r)$ parameters $\varepsilon=1$ and $\sigma=1$, and was simulated via molecular dynamics\cite{Anderson2020} in the NVT ensemble at $k_B T = 1.6$ and number density $\rho = 1.6$.
    (a) Snapshots show the system before, during, and after crystallization.
    (b) Corresponding distributions of $g(r)$ show a transition to sharp peaks.
    (c) Distributions of coordination numbers also show a transition toward a sharp peak at 12.
    (d) Similarly, the bond orientational order diagram shows a transition from a more isotropic distribution to one characterized by sharp peaks. 
    The orientation of the bond orientational order diagram does not correspond to the structure orientation shown in (a), and values in each bin of the bond orientational order diagram are clipped to the 95th percentile to aid visualization.
    For (c) and (d), neighbors of each particle were defined as those within the distance $r_{\max}=1.1$ from its center, which is the distance of the minimum after the first peak of $g(r)$. 
    All snapshots were rendered in \texttt{OVITO} and the analysis was performed via \texttt{freud}.
    }
    \label{fig:LJ1}
\end{figure}

By using only the three metrics we have described, which are the simplest characterizations of local structure, we can already detect crystallization and reveal information about order in liquid and crystal phases in many systems.
Here we use these three metrics to characterize a simple form of crystallization in a Lennard-Jones system.\cite{Schwerdtfeger2024}
This system, in which particles interact with each other with the pair potential $U(r) = 4\varepsilon \left[ (\sigma/r)^{12} - (\sigma/r)^{6} \right]$, is known to crystallize into the fcc and hcp structures in specific density and temperature regimes.\cite{Schwerdtfeger2024}
The parameters $\varepsilon$ and $\sigma$ set the energy and length scales, respectively, and $r$ represents the distance between the centers of any particle pair.
Figure~\ref{fig:LJ1}(a) shows three snapshots of a system prior to crystallization into the fcc structure, during crystallization, and after crystallization. 
Note the visual emergence of layers of ordered particles throughout the system.
This emerging order can also be seen in the transition to narrow and well-defined peaks in $g(r)$ [Fig.~\ref{fig:LJ1}(b)], a shift in the coordination number distribution toward a sharp peak at 12, which is the number of nearest neighbors per particle in the fcc structure [Fig.~\ref{fig:LJ1}(c)], and the evolution of the bond orientational order diagram from a more isotropic distribution to one with sharp peaks [Fig.~\ref{fig:LJ1}(d)].
The latter additionally indicates the presence of stacking faults in the system, because it contains peaks indicating multiple orientations of the fcc environment.
It is distinct from the bond orientational order diagram that would be calculated from a single fcc grain [Fig.~\ref{fig:bod}(b)].
The presence of these stacking faults will be discussed using more sophisticated characterization techniques in Sec.~\ref{section:stacking}.

\section{Harmonic order parameters}

Harmonic order parameters use \emph{harmonics} to characterize local environments.
For each environment, the angular information associated with each vector $\mathbf{r}_{ij}$ is collected into one signal, and this signal's spatial periodicity is analyzed by harmonic decomposition.
The spatial periodicity of the environment is related to its rotational symmetry, and so harmonic order parameters are used to quantify rotational symmetries of local particle environments in a fast, efficient manner.
For a detailed review of harmonic order parameters, see Ref.~\onlinecite{Keys2011}. 
We will first discuss the use of harmonic order parameters to characterize particle environments in 2D.
We will assume that every environment has $N$ vectors, for notational simplicity.

\subsection{Two dimensions}\label{section:2D}
The harmonic analysis of any 2D environment sheds light on its rotational symmetry by analyzing the periodicity of the set of angles \{$\theta_{ij}$\} that each environment vector makes with respect to a reference line.

To begin, we construct a probability density distribution describing the environment of particle $i$ from the angle set \{$\theta_{ij}$\}:
\begin{align}
    f_i(\theta) = \frac{1}{N} \sum_{j=1}^N \delta (\theta - \theta_{ij})\,,
\end{align}
where $\delta (\theta)$ is the Dirac delta function.
The probability density $f_i(\theta)$ can be integrated over any range of angles to give the probability of finding one of particle $i$'s neighbor bonds within that range.
This signal is one-dimensional and periodic in $2\pi$.

To perform the harmonic decomposition of the signal $f_i(\theta)$, we write it as the  Fourier decomposition 
\begin{align}
    f_i(\theta) = \frac{1}{2\pi} \sum_{\ell=-\infty}^{\infty} \psi_{\ell,i} e^{-i\ell\theta}\,,
\end{align}
where the $e^{-i\ell\theta}$ are periodic basis functions associated with each frequency $\ell$, and $\psi_{\ell,i}$ are the Fourier coefficients that quantify the projection of $f_i(\theta)$ onto each basis function (in other words, the strength of the contribution of each basis function to the overall signal).
We have chosen to normalize $f_i(\theta)$ by $2\pi$ and use the complex conjugate of the usual basis functions $e^{i\ell\theta}$ in the Fourier decomposition, so that the Fourier coefficients are more interpretable.
They will be used to analyze the rotational symmetry of the environment.

\begin{figure}[h]
    \centering
    \includegraphics[width=0.5\textwidth]{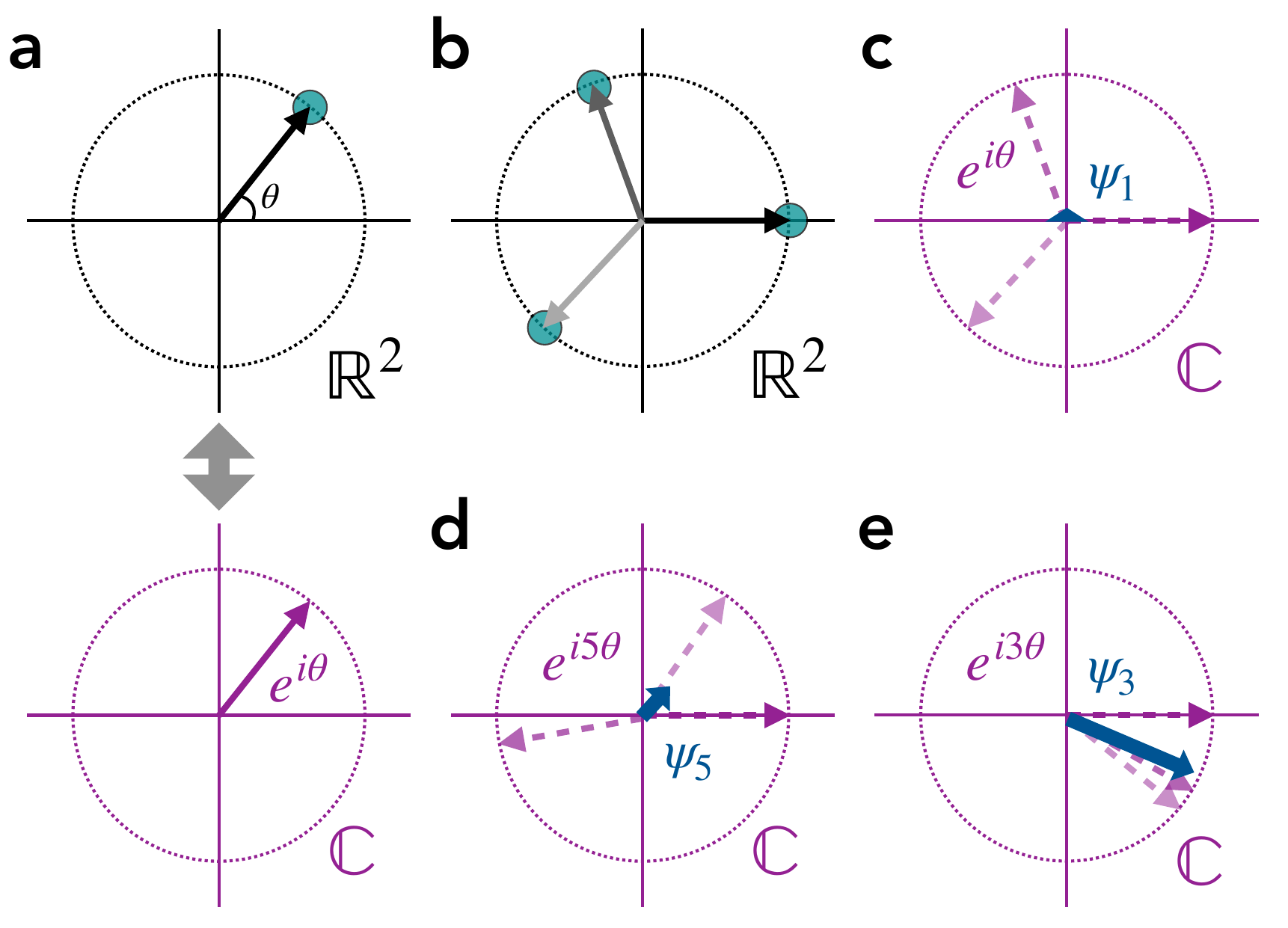} 
    \caption{
    Two-dimensional harmonic order parameters characterize point distributions on the unit circle.
    (a) The vector pointing from the origin to a point on the unit circle in $\mathbb{R}^2$ is equivalent to the vector pointing from the origin to the point $e^{i \theta}$ in $\mathbb{C}$.
    This equivalence is useful because it means that $\psi_\ell$ is the centroid of a pattern of angles on the unit circle if they are all multiplied by $\ell$ and drawn in $\mathbb{C}$.
    (b) An example pattern of angles on the unit circle, with almost three-fold rotational symmetry.
    (c) $\psi_1$ (the blue arrow) is the centroid of these points (pink arrows) in $\mathbb{C}$.
    Note that the centroid is almost at the origin, and thus the blue arrow has a very small magnitude.
    (d) $\psi_5$ (the blue arrow) is the centroid of the points formed from the multiplication of all angles by 5 (pink arrows) in $\mathbb{C}$.
    The pink arrows are not aligned, and thus the blue arrow has a relatively small magnitude.
    (e) $\psi_3$ (the blue arrow) is the centroid of the points formed from multiplication of all angles by 3 (pink arrows) in $\mathbb{C}$.
    The pink arrows are much more aligned, and thus the blue arrow has a large magnitude, approaching 1.
    }
    \label{fig:hex}
\end{figure}

The Fourier coefficients can be found using the orthogonality of the basis functions over the range $2\pi$ (see Appendix~\ref{section:psi1}):
\begin{align}
    \psi_{\ell,i} = \!\int_0^{2\pi} f_i(\theta) e^{i\ell\theta} d\theta\,.
\end{align}
We then substitute in the probability density form of $f_i(\theta)$:
\begin{subequations}
\begin{align}
    \psi_{\ell,i} & = \!\int_0^{2\pi} f_i(\theta) e^{i\ell\theta} d\theta  \\
    &= \! \int_0^{2\pi} \frac{1}{N} \sum_{j=1}^N \delta (\theta - \theta_{ij}) e^{i\ell\theta} d\theta  \\
    \psi_{\ell,i} &= \frac{1}{N} \sum_{j=1}^N e^{i\ell\theta_{ij}}\,.
\end{align}
\end{subequations}
$\psi_{\ell,i}$  maps a pattern of angles on the unit circle to a single complex number.
Each coefficient is intuitive if we consider it to be the centroid of the pattern when it is manipulated and drawn in the complex plane ($\mathbb{C}$) rather than the two-dimensional real plane ($\mathbb{R}^2$).
For example, to build $\psi_{1,i} = \frac{1}{N} \sum_{j=1}^N e^{i\theta_{ij}}$, we take the average of the complex numbers $e^{i\theta_{ij}}$ corresponding to the angles in the pattern.
Each number $e^{i\theta_{ij}}$ can be drawn as a vector in $\mathbb{C}$ with components $(\cos \theta_{ij}, i \sin \theta_{ij})$.
This vector in $\mathbb{R}^2$ corresponds to the unit vector in the pattern at angle $\theta_{ij}$ [Fig.~\ref{fig:hex}(a)].
Thus, $\psi_{1,i}$ can be thought of as the centroid of the pattern when it is drawn in $\mathbb{C}$ [see Figs.~\ref{fig:hex}(b) and (c)].
Generically, $\psi_{\ell,i}$ is the centroid of a pattern of angles if they are all multiplied by $\ell$ and drawn in the complex plane.
The magnitude of the centroids can be small or large  depending on the value of $\ell$ and the pattern itself [see Figs.~\ref{fig:hex}(d) and (e)].

The magnitude of $\psi_{\ell,i}$ is high if the pattern is approximately $\ell$-fold rotationally symmetric.
Graphically, this is because multiplying each angle in an $\ell$-fold rotationally symmetric set by $\ell$ collapses the set to a common angle and thus aligns all corresponding vectors in $\mathbb{C}$.
(Note that an $\ell$-fold rotationally symmetric set consists of $\ell$ angles that can be written as $\theta + 2\pi n /\ell$, where $n=0,1,\dots \ell-1$ and $\theta$ is the base angle of the set.
Multiplication of this set by $\ell$ collapses all angles to the angle $\ell \theta$ plus an integer multiple of $2\pi$.)
The alignment of the vectors in $\mathbb{C}$ maximizes the magnitude of their centroid and thus maximizes the magnitude of $\psi_{\ell,i}$ [see Fig.~\ref{fig:hex}(e)].
More generally, the magnitude of $\psi_{\ell,i}$ is maximal if the angle set consists of multiple subsets of $\ell$-fold rotationally symmetric angles (see Appendix~\ref{section:psi2}).

Since $\psi_{\ell,i}$ is the centroid of a pattern of angles in $\mathbb{C}$, it is not a rotationally invariant metric.
In other words, if the pattern rotates, its centroid in $\mathbb{C}$ also rotates, and thus $\psi_{\ell,i}$ changes.
However, $\vert \psi_{\ell,i} \vert$, the magnitude of the centroid in $\mathbb{C}$, does not change.
For that reason, it is common to  use
\begin{align}
    \vert \psi_{\ell,i} \vert = \frac{1}{N} \left\vert \sum_{j=1}^N e^{i\ell \theta_{ij} } \right\vert
\end{align}
as an order parameter to characterize the rotational symmetry of particle environments, because it is agnostic with respect to the orientation of each environment.

$\psi_\ell$ is used widely to investigate orientational order in 2D or quasi-2D particle systems.
For example, it has been used to examine phase transitions in experimental\cite{Walsh2016} and simulated\cite{Anderson2017} systems of anisotropic particles, the influence of surfaces on orientational ordering,\cite{Dullens2006} the influence of polydispersity on the structure of jammed systems of isotropic particles,\cite{Keim2015} the relation between structure and particle rearrangement under mechanical perturbation,\cite{Richard2020a} and the transition between liquid, hexatic, and solid phases in 2D.\cite{Marcus1996}

\begin{figure}[ht]
    \centering
    \includegraphics[width=0.5\textwidth]{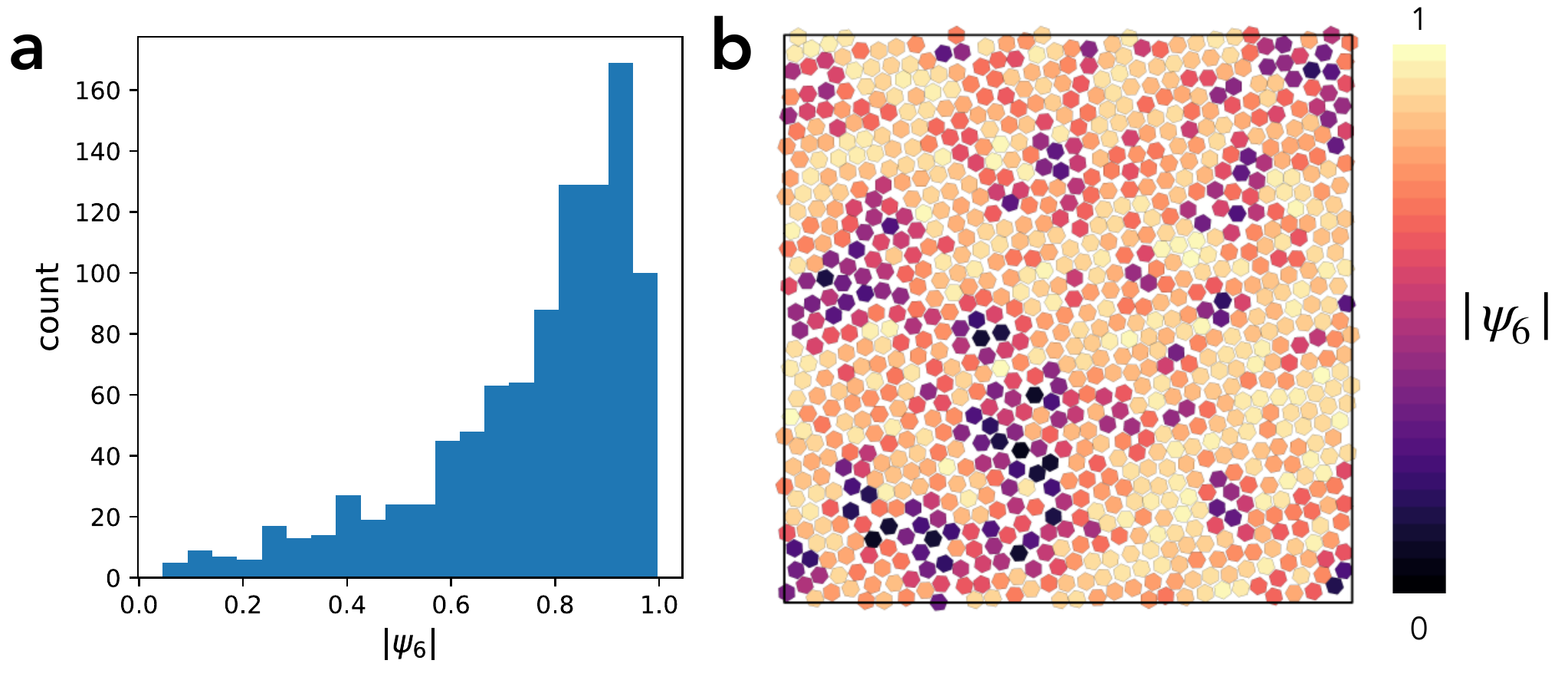} 
    \caption{
    Characterization of the 2D hexatic phase using $\psi_6$.
    (a) Histogram of $\vert \psi_6 \vert$ across all particles in a hard hexagon system at packing fraction $\phi=0.69$.
    The system was simulated via hard particle Monte Carlo sampling.\cite{Anderson2015a}
    Note the significant number of particles with $\vert \psi_6 \vert$ values approaching 1.
    (b) Image of the system with particles colored by $\vert \psi_6 \vert$.
    Note the large clusters of particles with high values of $\vert \psi_6 \vert$ that span the system; these are hallmarks of the long-range orientational order that characterizes the hexatic phase.
    To calculate $\psi_6$, the neighborhood of each particle was defined as its six nearest neighbors. 
    The system snapshot is rendered in \texttt{OVITO}, and analysis was performed via \texttt{freud}.
    }
    \label{fig:2d}
\end{figure}

Figure~\ref{fig:2d} shows the usefulness of $\psi_{\ell,i}$ to characterize orientational order in a 2D system of hard hexagons. 
This system is known to undergo a continuous phase transition from a fluid to a solid, passing through an intermediary hexatic phase with long-range six-fold orientational order and short-range positional order.\cite{Anderson2017} 
This phase transition is predicted by the so-called Kosterlitz-Thouless-Halperin-Nelson-Young theory of two-step melting\cite{Kosterlitz1973}, and precisely characterizing it in a range of systems is of general interest to the statistical physics community. 
The order parameter $\psi_{6}$ is often used for this purpose, because its magnitude $\vert \psi_{6} \vert$ reflects the degree of six-fold orientational order surrounding each particle.
In Fig.~\ref{fig:2d}, a system in the hexatic phase at packing fraction $\phi=0.69$ is shown.
Orientationally-ordered particle environments with high values of $\vert \psi_{6} \vert$ can be seen in large clusters throughout the system. 

\subsection{Three dimensions}

The harmonic analysis of the rotational symmetry of 3D environments follows a similar logic to the 2D case, but with the spherical harmonics replacing the $e^{i\ell \theta}$ harmonics as basis functions.
We first construct a probability density distribution to represent the environment on the surface of the unit sphere:
\begin{align}
    f_i(\theta, \phi) = \frac{1}{N} \sum_{j=1}^N \frac{1}{\sin \theta} \delta(\theta-\theta_{ij}) \delta(\phi-\phi_{ij})\,.
\end{align}
This distribution can be integrated over any solid angle to determine the probability of finding one of particle $i$'s neighbor bonds in that solid angle. 

We can then perform a harmonic decomposition of the signal $f_i(\theta, \phi)$ by writing it in multipole expansion form as
\begin{align}
    f_i(\theta, \phi) = \sum_{\ell=0}^{\infty} \sum_{m=-\ell}^{\ell} q_{\ell m,i} Y_{\ell m}^*(\theta, \phi)\,.
\end{align}
The spherical harmonics are defined as 
\begin{equation}
Y_{\ell m} \left( \theta, \phi \right) = (-1)^m \left[ \frac{(2\ell+1)(\ell-m)!}{4\pi (\ell+m)!} \right]^{1/2} P_{\ell m} \left( \cos\theta \right) e^{im\phi}\,.
\end{equation}
They are basis functions associated with each pair of indices $(\ell,m)$, and the $q_{\ell m,i}$ coefficients quantify the projection of $f_i(\theta, \phi)$ onto each basis function (in analogy to the Fourier coefficients in  2D).
We will use the $q_{\ell m,i}$ coefficients to quantify the rotational symmetry of the environment in direct analogy to  the 2D case.
(As in 2D, we actually use the complex conjugate of the basis functions to make the coefficients most easily interpretable.)

The coefficients can be found using the orthogonality of the spherical harmonics over the unit sphere (see Appendix~\ref{section:Q1}):
\begin{align}
    q_{\ell m,i} = \!\int_0^{2\pi} d\phi \int_0^{\pi} \sin \theta~d\theta f_i \left(\theta, \phi \right) Y_{\ell m} \left(\theta, \phi \right)\,.
\end{align}
We then substitute in the probability density form of $f_i(\theta, \phi)$ as follows:
\begin{subequations}
\begin{align}
    q_{\ell m,i} &= \!\int_0^{2\pi} d\phi \!\int_0^{\pi} \sin \theta~d\theta f_i \left(\theta, \phi \right) Y_{\ell m} \left(\theta, \phi \right)  \\
    &= \!\int_0^{2\pi} d\phi \int_0^{\pi} \sin \theta~d\theta \frac{1}{N} \sum_{j=1}^N \frac{1}{\sin \theta} \delta(\theta-\theta_{ij})  \notag \\
    &\qquad \times \delta(\phi-\phi_{ij}) Y_{\ell m} \left(\theta, \phi \right)  \\
    q_{\ell m,i} &= \frac{1}{N} \sum_{j=1}^N Y_{\ell m} (\theta_{ij}, \phi_{ij})\,.
\end{align}
\end{subequations}
For each value of $\ell$, there are $2\ell+1$ $q_{\ell m,i}$ coefficients corresponding to each allowed value of $m$.
Like the parameter $\psi_{\ell,i}$, these coefficients are not rotationally invariant, because the spherical harmonics themselves are not rotationally invariant.
In other words, if we consider the $(2\ell+1)$-dimensional vector $\mathbf{q}_{\ell,i}$, with each vector component corresponding to $q_{\ell m,i}$ for an allowed value of $m$, rotation scrambles its components.
However, its magnitude $\vert \mathbf{q}_{\ell,i} \vert = \sqrt{\sum_m \vert q_{\ell m,i} \vert^2}$ does not change under rotation, in analogy to the magnitude $\vert \psi_{\ell,i} \vert$.
Thus, the usual harmonic order parameter to characterize the rotational symmetry of particle environments in 3D is agnostic with respect to environment orientation and is defined as
\begin{equation}
Q_{\ell,i} \equiv \sqrt{\frac{4\pi}{2\ell+1} \sum_{m=-\ell}^\ell \vert q_{\ell m,i} \vert^2}\,.
\end{equation}
This number is also called a Steinhardt order parameter.\cite{Steinhardt1983}
The prefactor $4\pi/(2\ell+1)$ normalizes the parameter such that its maximum value is 1, regardless of the value of $\ell$ (see Appendix~\ref{section:Q2}).

\begin{figure}[ht!]
    \centering
    \includegraphics[width=0.5\textwidth]{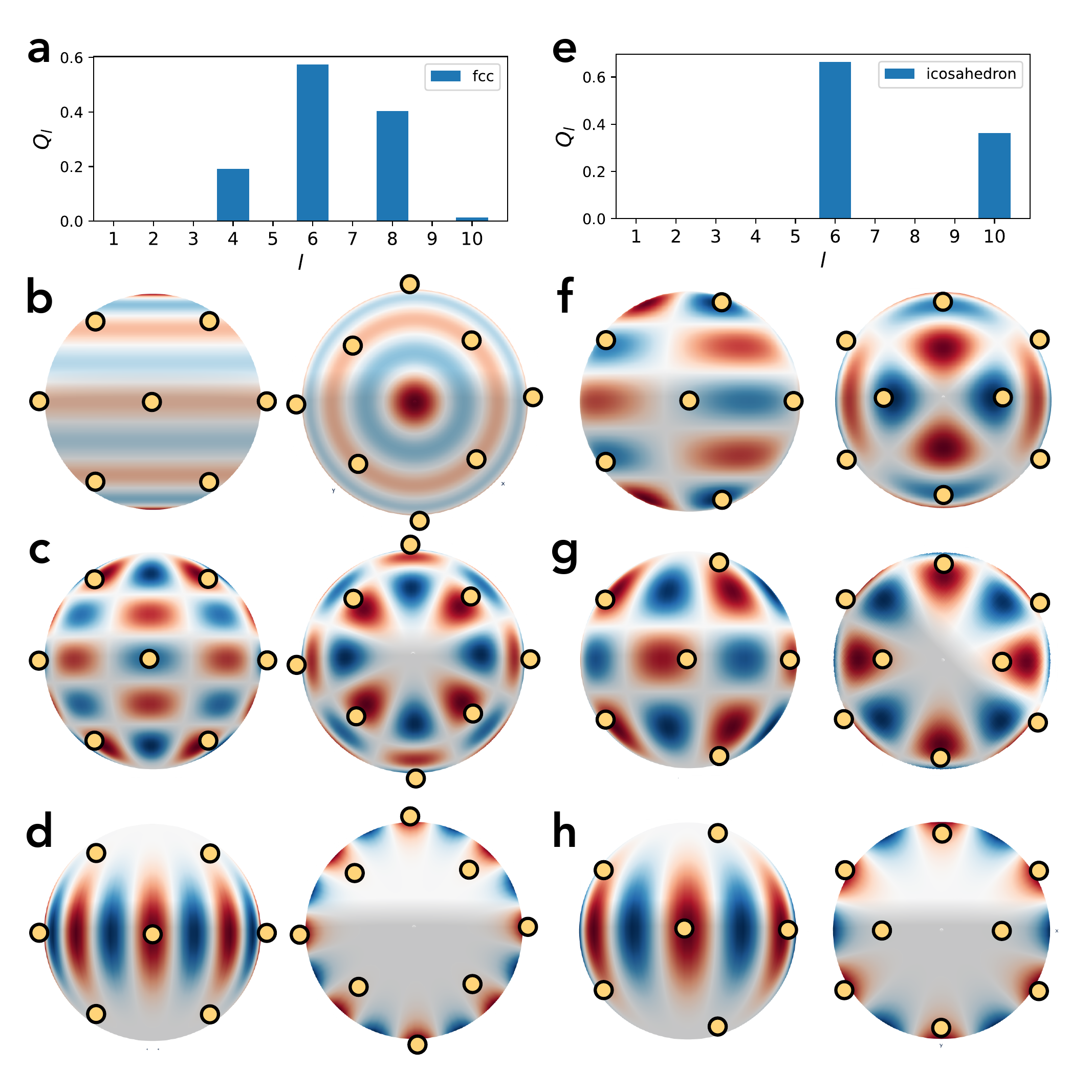} 
    \caption{
    3D harmonic order parameters characterize point distributions on the unit sphere.
    Shown are $Q_\ell$ parameters for $\ell \leq 10$ for point distributions corresponding to (a) the fcc and (e) icosahedral local environments.
    For any value of $\ell$, $Q_\ell$ is large if the environment overlaps significantly with a subset of the corresponding $2\ell+1$ spherical harmonics.
    (b)--(d) For the fcc environment, $Q_8$ is large because $\vert q_{8-8}\vert$, $\vert q_{8-4}\vert$, $\vert q_{80}\vert$, $\vert q_{84}\vert$, and $\vert q_{88}\vert$ are nonzero. 
    The fcc environment overlaps significantly and constructively with high-magnitude regions of the corresponding spherical harmonics.
    This overlap is shown for the spherical harmonics corresponding to the three highest values of $\vert q_{8m}\vert$: (b) $Y_{80}$, (c) $Y_{84}$, and (d) $Y_{88}$.
    (f)--(h) For the icosahedral environment, $Q_6$ is large because $\vert q_{6-6}\vert$, $\vert q_{6-4}\vert$, $\vert q_{6-2}\vert$, $\vert q_{60}\vert$, $\vert q_{62}\vert$, $\vert q_{64}\vert$, and $\vert q_{66}\vert$ are nonzero due to overlap with the corresponding spherical harmonics.
    These overlaps are shown for the spherical harmonics corresponding to the three highest values of $\vert q_{6m}\vert$: (f) $Y_{6-2}$, (g) $Y_{64}$, and (h) $Y_{66}$.
    The environment points are shown as yellow circles, and the spherical harmonics are represented as heat maps over the surface of a sphere: red regions correspond to positive values, blue regions to negative values, and white regions to zero. 
    Two different orientations of each environment/spherical harmonic are shown to more fully illustrate the overlap between them.
    We show the real part of the spherical harmonics only, because the environment orientations we have chosen mean that the environments do not overlap with the imaginary part of any of the spherical harmonics.
    }
    \label{fig:Ql}
\end{figure}

The Steinhardt order parameters, like their two-dimensional counterparts, provide information about the rotational symmetry of the environment.
However, unlike in the 2D case, there is no one-to-one relation between the values of $\ell$ and rotational symmetries.
Instead, $Q_\ell$ is large if the environment projects onto the subset of spherical harmonics of degree $\ell$ (and the associated allowed values of $m$) in a significant way.
Typically, environments are distinguished from each other using $Q_\ell$ parameters corresponding to several values of $\ell$.
We can consider an array of $Q_\ell$ parameters as a ``fingerprint" associated with a particular environment.
For example, if we consider all values of $\ell\leq 10$, the 12-particle fcc environment shown in Fig.~\ref{fig:envII}(a) has 2, 3, and 4-fold rotational symmetries, and especially high values of $Q_4$, $Q_6$, and $Q_8$ [Fig.~\ref{fig:Ql}(a)].
The 12-particle icosahedral environment shown in Fig.~\ref{fig:envII}(b) has 2, 3, and 5-fold rotational symmetries, and especially high values of $Q_6$ and $Q_{10}$ [Fig.~\ref{fig:Ql}(e)].
They can be distinguished from each other by their harmonic fingerprint, the vector of all $Q_\ell$ values for $\ell$ less than some maximum value.
Note that whether or not these $Q_\ell$ values are close to 1 does not indicate crystalline quality.
Rather, the set of values collectively-- the fingerprint-- indicates the specific rotational symmetry of the environment.

Figures~\ref{fig:Ql}(b)--(d) and \ref{fig:Ql}(f)--(h) show the projection of each environment onto three spherical harmonics associated with the highest values of $\vert q_{8m} \vert$ for the fcc environment and $\vert q_{6m} \vert$ for the icosahedral environment, respectively. 
We can see that each environment does not perfectly superimpose onto any spherical harmonic; rather, $\vert q_{\ell m} \vert$ is high if the environment and the spherical harmonic overlap significantly and constructively in regions of large (positive or negative) $Y_{\ell m}$.
Significant overlap happens if the rotational symmetry of the environment and the rotational symmetry of the spherical harmonic are compatible.
For the particular environment orientations we have chosen, Im$[q_{\ell m}] = 0$ for all $m$, so it is necessary only to visualize the projection of each environment onto the real part of the spherical harmonics of interest.
For generic orientations, $q_{\ell m}$ will have both real and imaginary parts. 
Also, environments project onto the spherical harmonics differently depending on their orientation, so values of $m$ associated with large $\vert q_{\ell m} \vert$ depend on the environment orientation.
However, regardless of environment orientation, $Q_\ell$ is unchanged.
Note that for both examples, $Q_\ell = 0$ for all odd values of $\ell$ because both environments have inversion symmetry (see Appendix~\ref{section:Q3}).

The following averaged form\cite{Lechner2008} of the Steinhardt order parameter is often used to characterize extended structure in particle systems:
\begin{align}
\bar{Q}_{\ell,i} &= \sqrt{\frac{4\pi}{2\ell+1} \sum_{m=-\ell}^\ell \vert \bar{q}_{\ell m,i} \vert^2} \\
\bar{q}_{\ell m,i} &= \frac{1}{N} \sum_{j=0}^{N} q_{\ell m,j}\,,
\end{align}
where the sum over $j$ runs over all $N$ neighbors of particle $i$ and particle $i$ itself.
The quantity $\bar{q}_{\ell m,i}$ thus includes orientational information averaged over all particle environments in the neighborhood of particle $i$.
The average $\bar{Q}_\ell$ is more stable under noise and characterizes structure on a longer length scale than $Q_\ell$, and is useful for examining the symmetries of crystal grains for which particles and their neighbors have similar local environments.

Steinhardt order parameters have been used to examine crystallization,\cite{Auer2004a, Lechner2008, Reinhardt2012} the glass transition and amorphous structure,\cite{Steinhardt1983, Leocmach2012, Tanaka2012, Yang2021c} and transitions between multiple crystal structures.\cite{Du2017, Wan2019} 
However, because the values of these order parameters are very sensitive to how the nearest neighborhood of each particle is defined,\cite{Mickel2013} caution must be taken when using Steinhardt order parameters to identify structure.

\section{Environment matching} \label{section:env}

The harmonic order parameters use only the angular information associated with each environment to characterize it, and thus omit significant information associated with bond lengths.
In this section we discuss a characterization method that incorporates that information by  ``environment matching,"  in which particle environments are compared with other nearby environments to determine their similarity.
Crystal structures can thus be identified by locating and analyzing similar particle environments.
This method has the advantage of being agnostic in its approach to structural characterization, because it does not seek to identify specific crystal structures, but rather identifies which particles are ``crystalline''  because they possess local environments that are repeated throughout the system.

Suppose that two environments are denoted as $\{ \bm{r}_{im} \}$ and $\{ \bm{r}_{jm'} \}$, where the first vector set represents the environment of particle $i$, the second set represents the environment of particle $j$, $m$ indexes over particle $i$'s $N$ neighbors, and $m'$ indexes over particle $j$'s $N$ neighbors.
We say that these two environments are similar enough to match if we can find a rotation $U$ and a one-to-one mapping between pairs of vectors in each set such that $\vert \bm{r}_{im} - U \bm{r}_{jm'} \vert < t$ for every mapping pair $(m,m')$ for some threshold $t$.
Matching can either be sensitive to rotation, in which case otherwise identical environments of different orientations are considered distinct, or rotationally invariant, in which case otherwise identical environments of different orientations are considered indistinct. 
If matching is sensitive to rotation, $U$ is set to the identity matrix. 
If matching is rotationally invariant, the method first attempts to solve the registration problem of finding the rotation $U$ that minimizes the root-mean-squared difference between the environments, then attempts to find the more restrictive pairwise mapping according to the threshold.

\begin{figure}[ht!]
    \centering
    \includegraphics[width=0.5\textwidth]{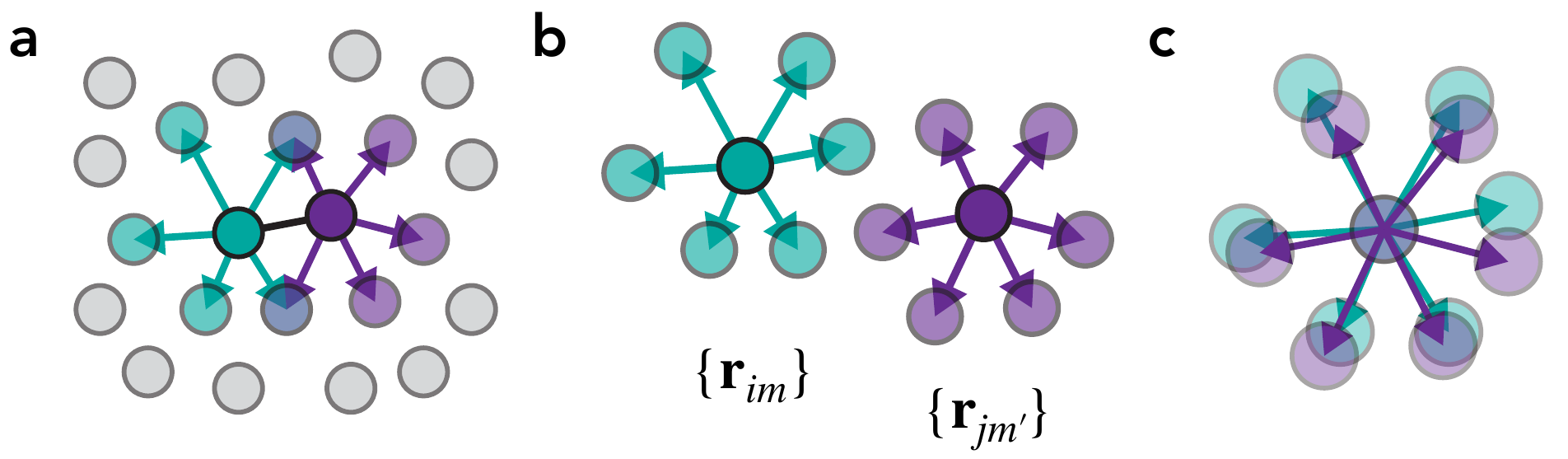} 
    \caption{
    Schematic of the environment-matching technique.
    (a) The environment of each particle is defined as the set of vectors pointing from the particle to its $N$ nearest neighbors.
    The environments of two neighboring particles are illustrated in green and purple and (b) denoted as $\{ \mathbf{r}_{im} \}$ and $\{ \mathbf{r}_{jm'} \}$.
    (c) Two environments match if there exists a one-to-one mapping between pairs of vectors in each set (and an optional rotation $U$) such that $\vert \bm{r}_{im} - U \bm{r}_{jm'} \vert < t$ for every mapping pair $(m,m')$.
    The threshold $t$ is chosen by the user.
    }
    \label{fig:env_match}
\end{figure}

The registration problem (finding $U$ and an appropriate one-to-one mapping) is nontrivial and of great interest to those concerned with image recognition. 
Half of this problem is solved by the Kabsch algorithm,\cite{Kabsch1976} which finds the optimal rotation to minimize the root-mean-squared difference between two \emph{labeled} sets of points centered about the origin.
In other words, each point in each set is distinguishable and labeled by its order in the set, and the root-mean-squared difference minimized by the Kabsch algorithm is taken over every pair of points with the same label in each set.
We derive the Kabsch algorithm in Appendix~\ref{section:kabsch}.
The real issue in finding the root-mean-squared difference between indistinguishable point sets is one of permutation. In principle, we could compare every permutation of one point set against the other point set and use the Kabsch algorithm to find the minimum root-mean-squared difference for all permutations. 
However, the number of permutations of $N$ points is $N!$, meaning that the number of calculations required to exhaustively solve the problem combinatorially is not practical.
In the following we describe a brute-force solution to the registration problem, implemented in the \texttt{Environment} module within the package \texttt{freud}, that works well in most cases.
\begin{enumerate}
\item 3 points are chosen at random from the set $\{\bm{r}_{jm'}\}$.
\item 3 points are chosen from the set $\{\bm{r}_{im}\}$.
The points are chosen in order from an exhaustive list of all possible 3-point permutations and combinations.
\item The matrix $U$ is found, which minimizes the root-mean-squared difference between these two subsets of 3 vectors each.
We use three vectors because many particle environments in a typical system are approximately misaligned by a rigid rotation.
\item The root-mean-squared difference between the full point set $\{\bm{r}_{jm'}\}$ and the full rotated point set $\{U\bm{r}_{im}\}$ is found. 
The root-mean-squared difference is computed over the pairing of points found by looping over each point in $\{\bm{r}_{jm'}\}$ and pairing it with the nearest point in $\{U\bm{r}_{im}\}$ that is not already matched to any other point in $\{\bm{r}_{jm'}\}$. 
(This method is not guaranteed to find the absolute minimal root-mean-squared difference; to do that, it is necessary to solve the assignment problem.\cite{Burkard2012})
\item Steps 2--4 are repeated either until the root-mean-squared difference between the full sets is less than the chosen tolerance $10^{-6}$, or until every possible combination and permutation of 3 points in $\{\bm{r}_{im}\}$ has been considered. 
\item The returned root-mean-squared difference (and optimal rotation and pairing) is the minimal one over all those calculated in step~5. 
\end{enumerate}
After the optimal rotation and pairing is found, the vector sets are subjected to a stricter criterion for matching necessitating that every optimally rotated pair must have a displacement between them that is below the threshold $t$.
This algorithm is computationally expensive.
Nevertheless, it can be very helpful in locating crystal grains throughout a system for which the crystal structure type is not known.

If the user opts to look for matching environments in a manner that is sensitive to rotation, it is not necessary to employ registration to match environments.
In that case, the algorithm implemented by the \texttt{Environment} module of \texttt{freud} is much simpler and faster: Each point in $\{\bm{r}_{jm'}\}$ is looped over, and paired with any unpaired point in $\{\bm{r}_{im}\}$ if the displacement between the points is below the threshold. 
If a complete 1-to-1 map is found in this way, then the environments match.

Which threshold to use is an important choice. 
If $t$ is small, the method uses a strict criterion for matching, and there is a risk that similar particle environments, differing only because of thermal or other noise, will not be detected. 
If $t$ is large, the method uses a loose criterion for matching, and quite different particle environments might spuriously register as matching.
A good rule of thumb that works well in most cases is to use values of the threshold ranging from $t=0.2 r_{\max}$ to $t=0.35 r_{\max}$, where $r_{\max}$ is the average nearest-neighbor distance in the system.
This range of values has been empirically found to  meaningfully distinguish particle environments in the presence of noise.

Environment matching has been used in multiple contexts, including the monitoring of crystallization in hard particle systems,\cite{Teich2019} the identification of crystal grains in bi-disperse jammed systems,\cite{Teich2021b} and the structural characterization of white matter in the human brain.\cite{Teich2021a}

\section{Structure identification}
The environment matching algorithm we have described  distinguishes between environments within a single system, but does not identify what those environments are. 
It is possible to use the same conceptual process to compare to a library of known local particle environments, and therefore identify crystal structures. 
There are several established methods for doing this, in addition to novel machine learning-based algorithms that seek to address limitations in the conventional techniques. 

Most structure identification techniques follow a similar algorithm: use some measure to quantify the local environment as a ``fingerprint'' of a particle, solve for rotational and scale invariance, and then compare to a limited set of candidate structures for which the fingerprint is already known. 
The drawback is that, because this method relies on the selection of candidate structures, it is most useful when identifying a simple, well-known structure, or in systems which are expected to crystallize into one of a handful of known options.

Structure identification also depends on the reliable, efficient calculation of a fingerprint for every particle. 
There are several common techniques for doing this. 
One technique is common neighbor analysis,\cite{Honeycutt1987,Faken1994} which uses a set $r_{\max}$ to identify ``bonds'' in a local neighborhood. 
This $r_{\max}$ is typically found using $g(r)$, as described in Sec.~\ref{section:rdf}. 
Common neighbor analysis uses the number of bonds to the reference particle as well as the topology of bonds between the set of neighbors to create the fingerprint, and compares them to a library. 
This method is typically robust for identifying the most common structures (simple cubic, bcc, fcc, etc.), but is not suited for all systems, \emph{e.g.}, diamond, because there are no bonds between the neighbors of a particle to use as an input.\cite{Maras2016}
Additionally, common neighbor analysis is sensitive to the value of $r_{\max}$, which limits the analysis of distorted or high-temperature systems.\cite{Larsen2016}
Extensions to common neighbor analysis use adaptive\cite{Stukowski2012} or user-identified bonds to overcome some of these problems, and consideration of second- or third-nearest-neighbors can expand the library of distinguishable structures.\cite{Maras2016} 
For a recent review of this method (and others), see Ref.~\citenum{Stukowski2012}.

In a system that is expected to be highly distorted, perhaps due to thermal fluctuations or strain, the $r_{\max}$-dependent approach that common neighbor analysis uses is noisy and a good alternative is polyhedral template matching.\cite{Larsen2016}
Rather than defining neighbors by $r_{\max}$, polyhedral template matching selects the nearest neighbors of each particle based on a combination of bond length and the Voronoi polyhedron, allowing for large fluctuations. 
To characterize the local neighborhood, polyhedral template matching takes the convex hull of the environment vectors, building a polyhedron which encloses every point in the set, but which contains no concave angles between facets.  This polyhedron can then be compared to a known library of template structures to identify the neighborhood structure.

To allow the comparison to be scale-invariant, the method converts the convex hull and each template in the library to graphs consisting of edges connecting vertices. 
Transforming the structures to graphs neglects all information about bond lengths and transforms the problem of comparing structures to determining whether they are identical by a graph isomorphism.
A small number of candidate templates in the library will have an identical graph to that of the neighborhood.
The polyhedral template matching method compares the actual structures of these templates to the neighborhood 
by rotating the neighborhood onto each template and selecting the template with the lowest mismatch. 
Polyhedral template matching has been used to identify structure in copper precipitate and copper platinum alloy simulations,\cite{Larsen2016} to monitor crystallization at solid interfaces,\cite{Yang2023} and to analyze 2D electron microscopy images,\cite{Britton2024} among other applications.


\section{Machine-learned metrics}
\label{section:additional}
To identify complex structures, machine-learned (ML) order parameters are being developed because they allow for many measures of the local or global environment to be input simultaneously.\cite{Martirossyan2024a}
ML-based techniques vary in both the architecture of the underlying classification process and the method of quantifying local structure. 

One technique proposed by Spellings et al.\cite{Spellings2018} uses bond orientational order to identify local structures by considering a range of nearest neighbors to produce a vector of many bond orientational order diagrams, providing detailed information about the local environment moving outward from the reference particle. 
This information produces a massive input dataset, which is difficult to categorize using conventional techniques, but which can be clustered by machine learning. 
The same model can be trained on known structures to identify a match with unknown systems, in a similar manner to the conventional algorithms. 
A benefit of this method is that it is capable of handling structures with multiple local environments, extending structure-matching to complex crystals.\cite{Spellings2018, Martirossyan2024}
Many other ML techniques have been developed to characterize local structure.
Recent studies have used a variety of techniques to classify local environments according to their crystal structure,\cite{Chung2022,Reinhart2017,Ziletti2018,Banik2023} understand rearrangement in disordered systems,\cite{Schoenholz2016} and monitor crystallization from fluids.\cite{Adorf2020} 
For a recent comprehensive review of the use of ML order parameters to characterize and design materials, see Ref.~\onlinecite{Martirossyan2024a}.

\section{Detecting stacking faults} \label{section:stacking}

Here we demonstrate how the metrics we have described can characterize in more detail the simple Lennard-Jones crystallization scenario discussed in Sec.~\ref{section:detectingcrystals}.
These metrics are capable of locating crystallizing regions of the system and stacking faults within these regions, in addition to signifying that crystallization occurs in general.

\begin{figure}[ht!]
    \centering
    \includegraphics[width=0.5\textwidth]{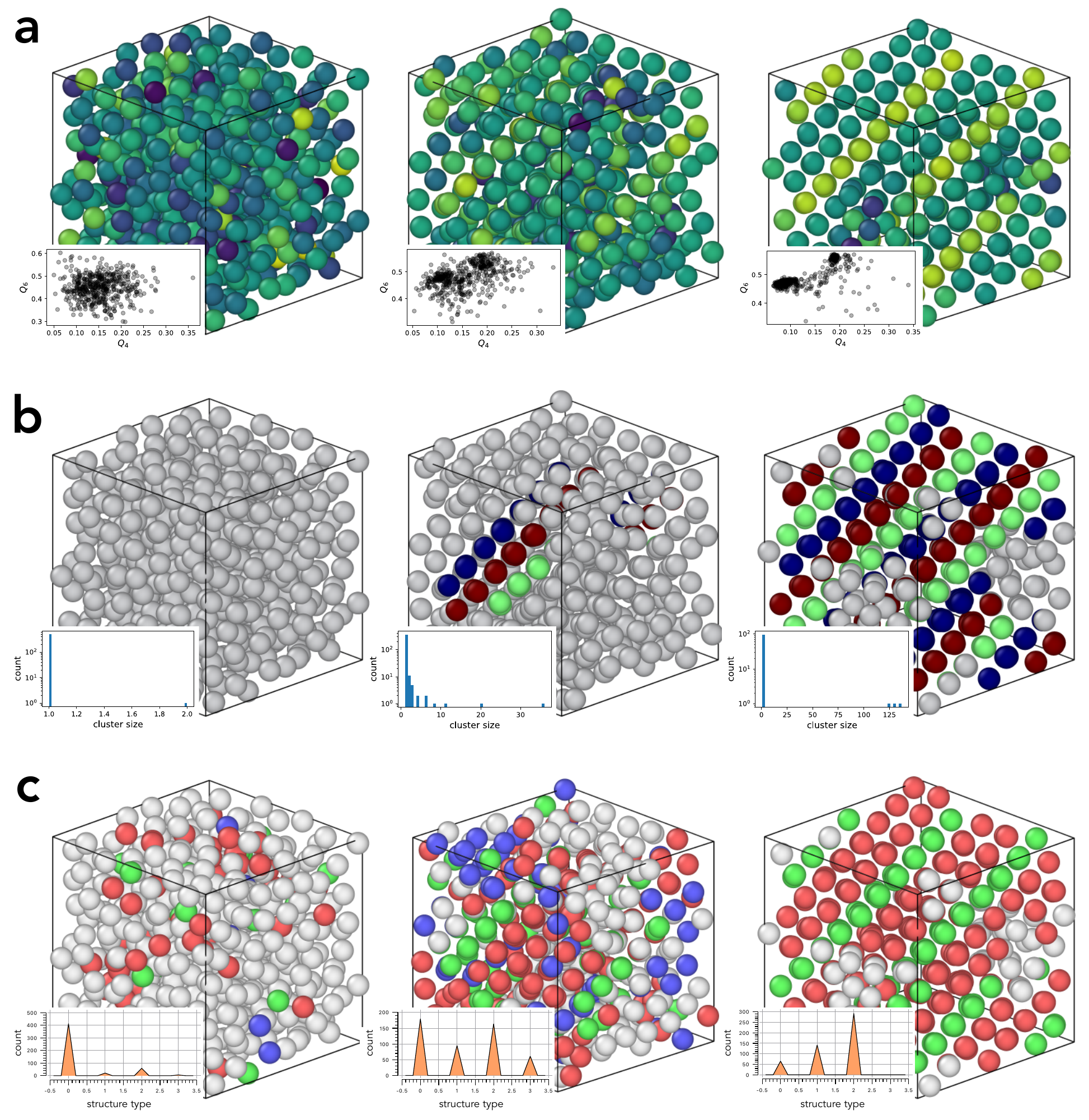} 
    \caption{
    Crystal grains and stacking faults in the Lennard-Jones system are located using Steinhardt order parameters, environment matching, and polyhedral template matching.
    The snapshots match those of Fig.~\ref{fig:LJ1} and show the system before, during, and after crystallization.
    (a) Particles are colored by $Q_6$, with purple corresponding to $Q_6=0.3$ and yellow corresponding to $Q_6=0.6$.
    Insets show the corresponding joint distributions of $Q_4$ and $Q_6$ for all particles.
    Note the transition to bands of particles with distinct values of $Q_4$ and $Q_6$ as crystallization proceeds.
    (b) Particles are colored according to crystal grain as determined by environment matching between all particle environments and the environments of their neighbors.
    (Colors may not be consistent from frame to frame, as clusters are colored according to an arbitrary cluster index.)
    The matching scheme has threshold $t=0.2$ and no registration.
    Particles are colored gray if their environments do not match the environments of their neighbors, or if they belong to crystal grains of size $\leq 10$.
    Inset images show the distributions of crystal grain size for each snapshot.
    Note the transition to three layers of distinct particle environments as crystallization proceeds.
    (c) Particles are colored according to polyhedral template matching with the root-mean-squared difference cutoff 0.15.
    The environment types shown are fcc (green, type 1), hcp (red, type 2), bcc (blue, type 3), and other (gray, type 0).
    Inset images show distributions of environment types for each snapshot.
    Polyhedral template matching confirms that the layers in the system have alternating hcp and fcc environments.
    Panels (a) and (b) use the same definition of particle neighborhoods as Fig.~\ref{fig:LJ1}. 
    Snapshots are rendered in \texttt{OVITO}, and analysis is performed via \texttt{OVITO} for polyhedral template matching and \texttt{freud} for Steinhardt order parameters and environment matching.
    }
    \label{fig:LJ2}
\end{figure}

Particle environments in the fcc structure have a distinct fingerprint of Steinhardt order parameters, as shown in Fig.~\ref{fig:Ql}(a).
Because we know that the Lennard-Jones system crystallizes into the fcc structure, we can judiciously choose a Steinhardt order parameter that is prominent in the fcc fingerprint to identify which particles crystallize into fcc in the simulation and which do not.
In Fig.~\ref{fig:LJ2}(a), particles in the three representative snapshots also shown in Fig.~\ref{fig:LJ1} (pre-crystallization, mid-crystallization, and post-crystallization) are colored by $Q_6$. 
Note the emergence of bands of particles with specific values of $Q_6$ as crystallization occurs.
Inset scatter plots show that particles not only have specific values of $Q_6$, but also $Q_4$, and that they largely fall into two clusters in $Q_4$--$Q_6$ space.
Values of $Q_4$ and $Q_6$ that correspond to these clusters roughly correspond to the fcc and hcp local environments.\cite{Steinhardt1983}
The hcp structure is distinct from fcc and features a different pattern of closely-packed hexagonal layers.
Thus, $Q_4$ and $Q_6$ indicate that the system self-assembles into a structure featuring bands of fcc and hcp due to stacking faults throughout the system.
$Q_4$--$Q_6$ maps are popular for their use in  understanding crystallization in various systems (see Fig.~2a in Ref.~\onlinecite{Leocmach2012}, for example).

Environment matching provides additional detail and corroborates the story told by $Q_6$ and $Q_4$.
In this case, environment matching can agnostically detect crystal grains without relying on any prior knowledge pertaining to fcc or hcp structures. 
Using the technique without registration identifies crystal grains of different orientations [Fig.~\ref{fig:LJ2}(b)].
We see that three layers exist in the system, colored green, blue, and red respectively, and they span the system due to periodic boundary conditions. 
The red layer corresponds to that with fcc-like Steinhardt order parameters, and the green and blue layers correspond to those with hcp-like Steinhardt order parameters.
In this analysis, each hcp-like layer is distinct, because each layer has a uniquely-oriented local environment.
(In contrast, because the Steinhardt order parameters are rotationally invariant, the hcp-like layers appear identical in Fig~\ref{fig:LJ2}(a).)
Histograms inset in each image show the distribution of cluster sizes identified by the matching process, illustrating the emergence of three significant clusters   during crystallization.

Polyhedral template matching confirms that the local environments in the crystallized system are those of the hcp and fcc structures [Fig.~\ref{fig:LJ2}(c)].
Histograms inset in each image show the count of each environment type, and illustrate the predominance of hcp environments throughout the system in the final crystal structure.
When the system is still crystallizing, bcc, fcc, and hcp environments are scattered throughout the system, although only some of them form crystal grains of more than 10 particles according to environment matching [Fig.~\ref{fig:LJ2}(b)].

\section{Detecting a transition between crystals}

\begin{figure}[ht!]
    \centering
    \includegraphics[width=0.5\textwidth]{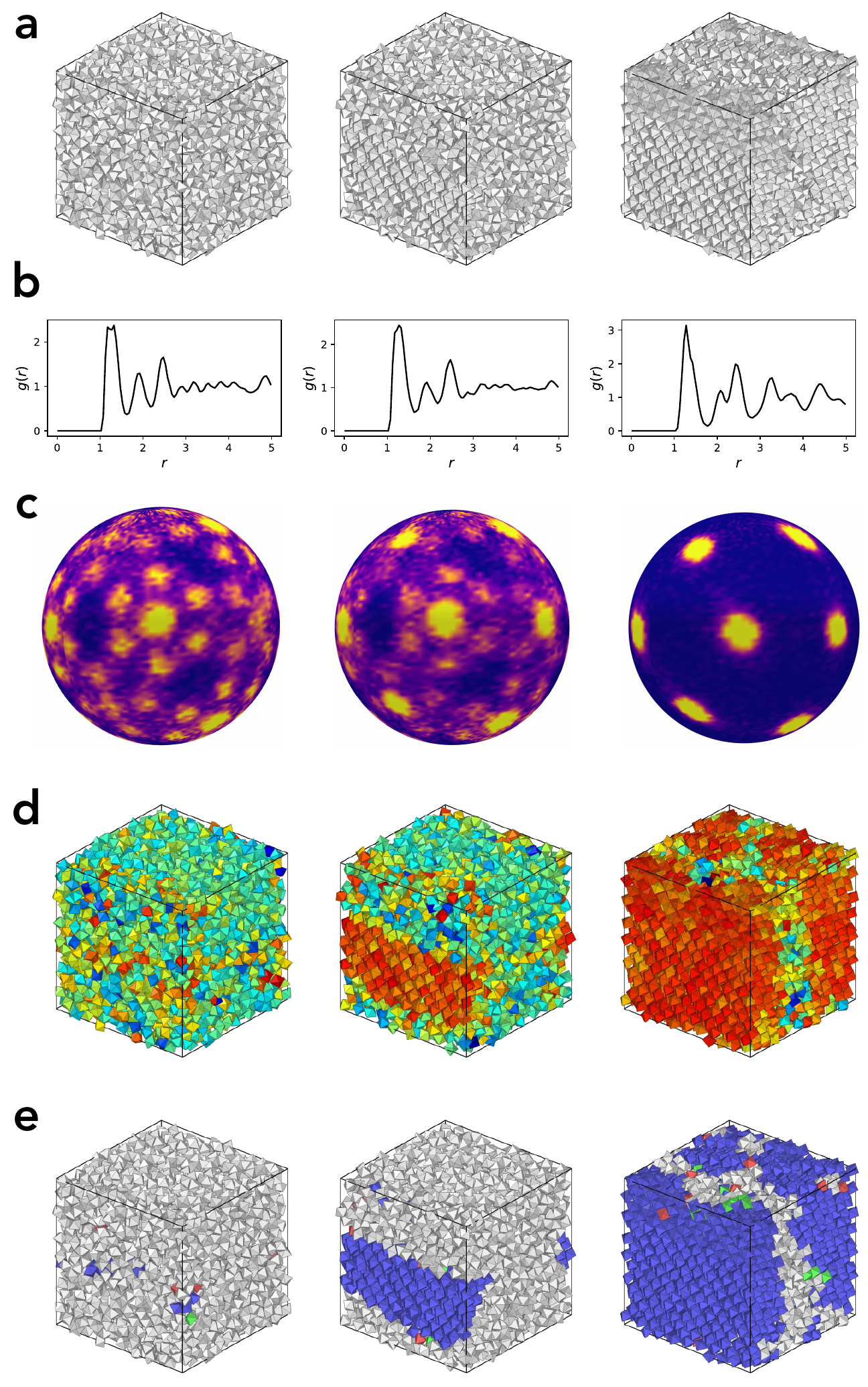}
    \caption{
    The Li-to-bcc structural transition in a system of hard truncated octahedra is partially captured by Steinhardt order parameters and polyhedral template matching.
    The system is simulated via hard particle Monte Carlo sampling at packing fraction $\phi=0.61$.
    (a) Snapshots show the system predominantly in the Li structure before the transition (left) and predominantly in the bcc structure after the transition (right).
    During the transition the system has elements of both structures (middle).
    (b) $g(r)$ distributions and (c) bond-orientational order diagrams corresponding to each snapshot show a clear transition from one crystalline structure to another.
    However, (d) the Steinhardt order parameter $Q_6$ and (e) polyhedral template matching only locate the emerging bcc crystal, and do not register the Li structure.
    $Q_6$ values for each particle in (d) range from 0.18 (blue) to greater than 0.55 (red). 
    The neighbors of each particle are all those within the distance $r_{\max}=1.7$ from its center: This distance is approximately the distance to the minimum after the first peak of $g(r)$ for both structures.
    The color scheme in (e) is the same as that of Fig.~\ref{fig:LJ2}(c), with blue indicating bcc environments. 
    The root-mean-squared difference cutoff is 0.15.
    }
    \label{fig:Li1}
\end{figure}

Figure~\ref{fig:Li1} shows the utility of the structural metrics we have introduced in a more complicated crystallization example.
In this case, a system of hard truncated octahedral particles undergoes a structural transition from a high-pressure lithium-like (Li) structure [Fig.~\ref{fig:Li1}(a)], left) to a bcc structure [Fig.~\ref{fig:Li1}(a)], right).
This data was previously published in Ref.~\citenum{Teich2019}.
Corresponding $g(r)$ distributions [Fig.~\ref{fig:Li1}(b)] and bond-orientational order diagrams [Fig.~\ref{fig:Li1}(c)] clearly indicate differences in these structures, as well as a cross-over structure featuring elements of both Li and bcc.
However, the Li structure cannot be detected by the Steinhardt parameter $Q_6$ [Fig.~\ref{fig:Li1}(d)] or polyhedral template matching [Fig.~\ref{fig:Li1}(e)].
These metrics clearly show the emergence of the bcc structure, characterized by larger $Q_6$ values and the bcc template environment, but they do not show any clear indications of structure in the left-most panels (corresponding to the Li structure).

\begin{figure}[ht!]
    \centering
    \includegraphics[width=0.5\textwidth]{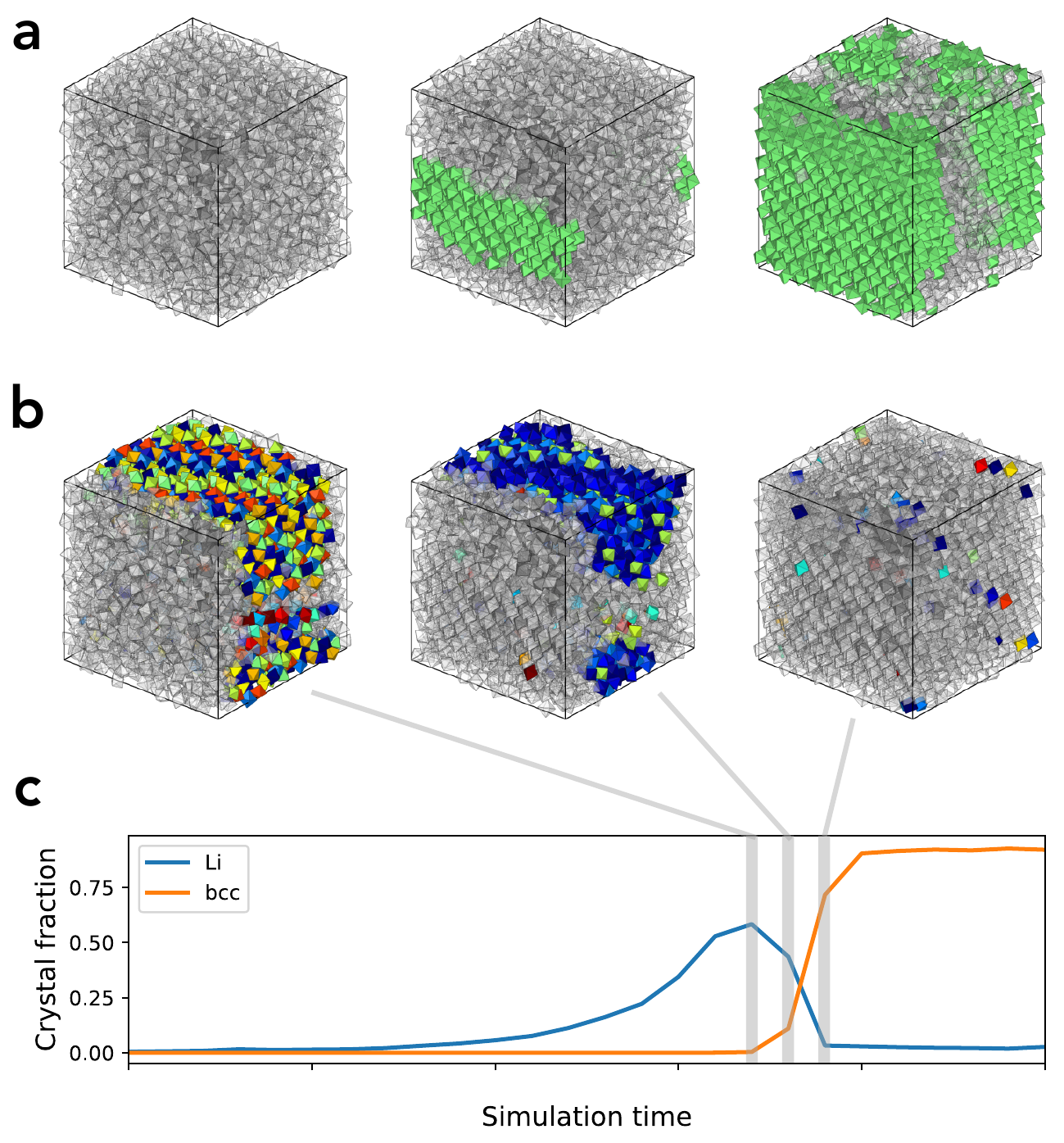}
    \caption{
    The Li-to-bcc structural transition in a system of hard truncated octahedra is fully captured by environment matching.
    All snapshots match those of Fig.~\ref{fig:Li1}, depicting the system undergoing a transition from Li to bcc.
    (a) The emerging bcc structure is located by an environment matching scheme in which environments are defined using the 14 nearest neighbors of every particle, and environment matching takes place between all particle environments and the environments of their 14 nearest neighbors.
    (b) The Li structure can also be located by an environment matching scheme in which environments are defined using the 11 nearest neighbors of every particle, and environment matching takes place between all particle environments and the environments of their 80 nearest neighbors.
    In panels a and b, particles are colored by crystal grain, and any particles in a crystal grain of size $\leq 5$ are transparent and gray.
    Note that the Li grain consists of eight particle colors, corresponding to eight unique orientations of one characteristic particle environment.
    (c) Environment matching quantifies the broader structural transition from Li to bcc over the course of the simulation.
    The plot displays the fraction of particles in each structure over a representative simulation time scale.
    In this case, the structure of particles is said to be  bcc or Li if they belong to a crystal grain of size $\geq 1$.
    The threshold for environment matching is $t=0.34$.
    }
    \label{fig:Li2}
\end{figure}

To locate the Li structure, we can employ the environment matching technique, which can be more informative for complicated crystal structures.
This technique is clearly capable of detecting the bcc particle environment, if environments are defined using the 14 nearest neighbors of every particle, and environment matching takes place between all particle environments and the environments of their 14 nearest neighbors [Fig.~\ref{fig:Li2}(a)].
Moreover, this technique can also detect the Li structure, if environments are defined using the 11 nearest neighbors of every particle, and environment matching takes place between all particle environments and the environments of their 80 nearest neighbors [Fig.~\ref{fig:Li2}(b)].
We choose 11 and 14 nearest neighbors to define environments in the Li and bcc structures, respectively, because these are the characteristic numbers of nearest neighbors in the first neighbor shell of the respective crystal structures.
We define the nearest neighbor shell of any particle as all particles within distance $r_{\max}=1.7$, which is the minimum after the first peak of each $g(r)$ in Fig.~\ref{fig:Li1}(b). 
The bcc structure is relatively simple, so the algorithm only needs to search for matches between the environments of every particle and its 14 nearest neighbors to identify bcc crystal grains.
However, to identify Li crystal grains, the field over which the algorithm searches for matching environments must be much larger (80, beyond the third nearest-neighbor shell) because the Li unit cell is bigger and thus particles with similar environments are further away from each other.
Additionally, particle pairs whose environments match according to the Li query parameters may also match according to the bcc query parameters, and care must be taken to ensure the Li and bcc particle subgroups are distinct.
To that end, we only categorize particles as members of the Li crystal structure if they are not also members of the bcc structure.
Using this definition of Li-like particles, we can see a significant Li crystal grain in the system that disappears as the bcc crystal grain emerges.
We can even track the structural transition between Li and bcc over the course of the simulation, revealing the slow growth of the Li structure and its sharp decay as the system rapidly transitions to the bcc structure [Fig.~\ref{fig:Li2}(c)].

System snapshots in Figs.~\ref{fig:Li1} and \ref{fig:Li2} are rendered in \texttt{OVITO}, and analysis is performed via OVITO (for polyhedral template matching) and \texttt{freud} (for $g(r)$ distributions, bond orientational order diagrams, Steinhardt order parameters, and environment matching).

\section{Summary}
We have discussed in detail several important and common techniques for the characterization of local structure in particle systems.
However, there exist many other structural characterization methods, and additional reviews can be found in Refs.~\onlinecite{Stukowski2012} and \onlinecite{Tanaka2019}.
All techniques range widely in their complexity and the level of detail they provide, and it is common to use combinations of these techniques to paint as full a picture as possible of system structure.
We provided several example applications in this spirit, using a combination of structural parameters to investigate simple crystallization from a fluid and orientational ordering in a hexatic system.
All of the discussed methods are widely available in multiple software packages and we encourage interested readers to dive in and begin experimenting with structural characterization in their own systems---they will find it a rich, rewarding, and beautiful world.

\section{Suggested Problems}

\begin{enumerate}
    \item In systems with multiple crystalline grains, the local environments differ. 
    Consider a simple cubic system and sketch the bond orientational order diagram for (a) an ideal single crystal, (b) two grains oriented at an angle of $45^\circ$ with respect to each other, and (c) a system of many grains with thin-film texture, meaning that most grains have a single axis $\hat{z}$ aligned in the same direction, but the perpendicular directions $\hat{x}$ and $\hat{y}$ have no preferred orientation.

    \item Many order parameters are commonly presented as a histogram, including $g(r)$, the bond orientational order diagram, and the Steinhardt order parameters. 
    The information that can be extracted from a histogram is sensitive to the bin width. 
    To explore this sensitivity, generate a system consisting of points on a simple cubic lattice a distance $a$ apart, disturbed by Gaussian noise with standard deviation $0.05a$, and compute $g(r)$ using multiple bin widths. 
    Choose values between $a/100$ and $a/5$. 
    Observe how the apparent characteristics of the system change as the bin width increases and how the total compute time changes. 
    What features appear if the bin width is very small? 
    If it is large? 
    How does the ideal bin width and compute time change with system size?
    
    \item In simulations, boundary conditions impose a difficulty to calculations that is not typically a problem for experimental systems. 
    For the system generated in Problem~2, compute $g(r)$ with and without periodic boundary conditions. 
    What changes in $g(r)$? 
    
    \item Unlike other measures we have presented, $g(r)$ can be computed over distances that are $5$ to $10$ times larger than the minimum interparticle distance. 
    Consider a system bounded by a cubic box with periodic boundary conditions. 
    What is the maximum distance $r$ over which $g(r)$ is meaningful?

    \item In some systems, \textit{positional} and \textit{orientational} order arise at different densities or temperatures. 
    Phase transitions from fluid to solid can pass through an intermediary phase featuring long-range orientational order and short-range positional order, like the hexatic phase shown in Fig.~\ref{fig:2d}.
    To explore this phenomenon further, initialize a disordered 2D system of hard hexagons at packing fraction $\phi = 0.5$, and compress it slowly to packing fraction $\phi = 0.8$. 
    During this compression, calculate and monitor three order parameters: $g(r)$, $\psi_6$ for each particle, and the bond orientational order diagram using the orientation of each particle as the reference for calculating angles.
    The first parameter indicates global positional order, while the second and third parameters indicate local bond orientational order.
    How does the development of positional order in the system compare to the development of orientational order?
    Can you locate a range of densities over which the hexatic phase is stable?
    At what densities are the fluid and solid phases stable?
    Compare your results to those reported in Ref.~\onlinecite{Anderson2017}.

    \item Use the parameters discussed in Sec.~\ref{section:detectingcrystals} to simulate crystallization in a Lennard-Jones system. 
    For this system, the potential parameters are $\varepsilon=1$ and $\sigma=1$, and the thermodynamic parameters are $k_B T = 1.6$ and number density $\rho = 1.6$. 
    Initialize the system at low $\rho = 0.5$, and compress it slowly to induce crystallization. 
    Try different choices of which particles constitute a local environment: the $N$ nearest neighbors, those within an $r_{\max}$ chosen as the location of the first peak of  $g(r)$, and those within an $r_{\max}$ chosen as the location of the minimum after the first peak of  $g(r)$. 
    Observe how these definitions influence the evolution during crystallization of coordination number distributions, Steinhardt order parameter distributions, and crystal grains found via environment matching and polyhedral template matching.

    \item Simulate a disordered system at low temperature using a Kob--Andersen mixture.\cite{Kob1994, Kob1997}
    This model consists of two particle types, A and B, that interact via Lennard-Jones potentials. 
    For the A-A interaction, $\epsilon_{AA} = 1.0, \sigma_{AA} = 1.0$; for the B-B interaction, $\epsilon_{BB} = 0.5, \sigma_{BB} = 0.88$; and for the A-B interaction, $\epsilon_{AB} = 1.5, \sigma_{AB} = 0.8$. 
    The mixture consists of 80\% A particles and 20\% B particles.
    Calculate $g_{AA}(r)$, $g_{BB}(r)$,  and $g_{AB}(r)$.
    How are the bonding lengths different among particle types?
    Calculate bond orientational order diagrams and a set of Steinhardt order parameters for each particle type. 
    Are they different?

    \item Simulate a system of hard octahedra at packing fraction $\phi = 0.6$, well above the crystallization point.\cite{Cadotte2016} 
    Identify crystal grains throughout the system using environment matching. 
    What parameters are required to identify these grains?
    Identify the type of crystal structure using polyhedral template matching and/or Steinhardt order parameters.
    What crystal structure do the octahedra self-assemble into?
    Compare your results to those reported in Ref.~\onlinecite{Cadotte2016}.

\end{enumerate}

\section{Citation diversity statement}
Recent work has identified a bias in citation practices such that papers from women and other marginalized scholars in STEM are under-cited relative to expected rates. 
Here we sought to proactively consider choosing references that reflect the diversity of the field in thought, form of contribution, gender, and other factors. 
We use databases that store the probability of a name being carried by people of different genders to mitigate our own citation bias at the intersection of name and identity. 
By this measure (and excluding self-citations to the first and last authors of our current paper, and papers whose authors' first names could not be determined), our references contain 9.46\% woman(first author)/woman(last author), 5.41\% woman/man, 22.97\% man/woman, and 62.16\% man/man categorization. 
This method is limited in that names, pronouns, and social media profiles used to construct the databases may not, in every case, be indicative of gender identity. 
Furthermore, probabilistic studies of names cannot be used to detect citation costs that are specific to intersex, non-binary, or transgender people who are out to a large number of their colleagues. 
We look forward to future work that could help us to better understand how to support equitable practices in science.

\section{Acknowledgments}
We thank Julia Dshemuchadse, Chrisy Xiyu Du, Sylvie Shaya, Kamakshi Subramanian, and Greg van Anders for helpful discussions.
This work was supported by the International Human Frontier Science Program Organization (HFSPO) under grant RGEC33/2024.

\appendix
\section{Derivation of $\psi_\ell$} \label{section:psi1}
To find the Fourier coefficients $\psi_\ell$, we use the fact that the basis functions are orthogonal over the range $2\pi$:
\begin{align}
    \int_0^{2\pi} e^{-i\ell \theta} \left[ e^{-i\ell '\theta} \right]^* d\theta &= 2\pi \delta_{\ell,\ell'}, \label{ortho}
\end{align}
where the Kroenecker delta $\delta_{\ell,\ell'} = 1$ if $\ell=\ell'$ and is 0 otherwise.
We can use Eq.~\eqref{ortho} find the Fourier coefficients as follows:
\begin{subequations}
\begin{align}
    f_i(\theta) &= \frac{1}{2\pi}\sum_{\ell=-\infty}^{\infty}\psi_{\ell,i} e^{-i\ell \theta} \\
    \! \int_0^{2\pi} f_i(\theta) e^{i\ell '\theta} d\theta &= \frac{1}{2\pi} \! \int_0^{2\pi}\sum_{\ell=-\infty}^{\infty} \psi_{\ell,i} e^{-i\ell \theta} e^{i\ell'\theta} d\theta  \\
    &= \frac{1}{2\pi} \sum_{\ell=-\infty}^{\infty} \psi_{\ell,i} 2\pi \delta_{\ell,\ell'}  = \psi_{\ell',i} \, .
\end{align}
\end{subequations}

\section{$\psi_\ell$ and rotational symmetry} \label{section:psi2}
We derive that the magnitude of $\psi_{\ell,i}$ is high  if the pattern of $N$ neighbor angles surrounding particle $i$ is $\ell$-fold rotationally symmetric.
Suppose first that the pattern is $m$-fold rotationally symmetric, which means that the set of angles can be grouped into $m$ subsets, where each consecutive subset is rotated by $2\pi/m$ with respect to the previous subset.
Each subset consists of $N/m$ angles, because there are $N$ total angles in the set.
Thus, we can write each angle as $\theta_k + 2\pi n/m$, where $n = 0, 1,\dots m-1$ and $k = 1, 2, \dots N/m$.
$\theta_k$ denotes the value of the $k$th angle in the first subset.
Then we can write $\psi_{\ell,i}$ as
\begin{align}
    \psi_{\ell,i} = \frac{1}{N} \sum_{k=1}^{N/m} \sum_{n=0}^{m-1} e^{i\ell [\theta_k + \frac{2\pi n}{m}]}
\end{align}
Its squared magnitude is
\begin{subequations}
\begin{align}
    \vert \psi_{\ell,i} \vert^2 &= \frac{1}{N^2} \sum_{k=1}^{N/m} \sum_{k'=1}^{N/m} \sum_{n=0}^{m-1} \sum_{n'=0}^{m-1} e^{i\ell [\theta_k + \frac{2\pi n}{m}]} e^{-i\ell [\theta_{k'} + \frac{2\pi n'}{m}]}  \\
    &= \frac{1}{N^2} \sum_{k=1}^{N/m} \sum_{k'=1}^{N/m} \sum_{n=0}^{m-1} \sum_{n'=0}^{m-1} e^{i\ell [\theta_k - \theta_{k'}]} e^{i\ell \frac{2\pi (n-n')}{m}} \\
    &= \frac{1}{N^2} \left[ \frac{N}{m} + 2 \sum_{k=1}^{N/m} \sum_{k'>k} \cos[\ell(\theta_{k} - \theta_{k'})] \right]  \nonumber \\ 
    & \quad \times \left[ m + 2 \sum_{n=0}^{m-1} \sum_{n'>n} \cos \left[\frac{2\pi \ell}{m}(n - n') \right] \right]. \label{eq:last}
\end{align}
\end{subequations}
To maximize the squared magnitude, irrespective of the specific angle set, we  choose $\ell$ to maximize the last term on the right-hand side of Eq.~\eqref{eq:last}.
This term  does not depend on the angles themselves and is maximized when $\ell/m$ is an integer, so that the argument to the cosine is an integer multiple of $2\pi$ and thus the cosine has maximal magnitude.
Thus, $\vert \psi_{\ell,i} \vert^2$ is maximal when $\ell=m, 2m, 3m, \dots$\,.

\section{Derivation of $q_{\ell m}$} \label{section:Q1}
To find the coefficients $q_{\ell m}$, we use the orthogonality of the spherical harmonics over the surface of the unit sphere. 
The orthogonality relation is
\begin{equation}
\int_0^{2\pi} d\phi \int_0^{\pi} \sin \theta~d\theta  ~Y_{\ell m} \left( \theta, \phi \right) Y_{\ell'm'}^* \left(\theta, \phi \right) = \delta_{\ell\ell '} \delta_{mm'} .
\end{equation}
Then
\begin{subequations}
\begin{align}
&f_i \left(\theta, \phi \right) = \sum_{\ell=0}^\infty \sum_{m=-\ell}^{\ell} q_{\ell m,i} Y_{\ell m}^* \left( \theta, \phi\right) \\
&\int_0^{2\pi} d\phi \int_0^{\pi} \sin \theta~d\theta f_i \left(\theta, \phi \right) Y_{\ell'm'} \left(\theta, \phi \right) \nonumber \\
&= \int_0^{2\pi} d\phi \int_0^{\pi} \sin \theta~d\theta \sum_{\ell=0}^\infty \sum_{m=-\ell}^{\ell} q_{\ell m,i} Y_{\ell m}^* \left( \theta, \phi \right) Y_{\ell'm'} \left( \theta, \phi \right) \\
&= \sum_{\ell=0}^\infty \sum_{m=-\ell}^\ell q_{\ell m,i} \delta_{\ell \ell '} \delta_{mm'} = q_{\ell'm',i} \, .
\end{align}
\end{subequations}

\section{Normalization of $Q_{\ell}$} \label{section:Q2}
We show that the maximum value of $Q_{\ell,i}$ is  1, regardless of the environment $i$ and the value of $\ell$.
First, we replace the notation of $Y_{\ell m}(\theta_j, \phi_j)$ by $Y_{\ell m}(\hat{j})$, where $\hat{j}$ is the unit vector corresponding to the angle pair $(\theta_j, \phi_j)$. 
Then we note that
\begin{subequations}
\begin{align}
    Q_\ell ^2 &= \frac{4\pi}{2\ell+1} \sum_{m=-\ell}^\ell \vert q_{\ell m} \vert^2  \\
    &= \frac{4\pi}{2\ell+1} \sum_{m=-\ell}^\ell \left[ \frac{1}{N} \sum_{j=1}^N Y_{\ell m}(\hat{j}) \right] \left[ \frac{1}{N} \sum_{j'=1}^N Y_{\ell m}^*(\hat{j'}) \right]  \\
    &= \frac{4\pi}{2\ell+1} \frac{1}{N^2} \sum_{j=1}^N \sum_{j'=1}^N \sum_{m=-\ell}^\ell Y_{\ell m}(\hat{j}) Y_{\ell m}^*(\hat{j'})  \label{eq:third} \\
    &= \frac{4\pi}{2\ell+1} \frac{1}{N^2} \sum_{j=1}^N \sum_{j'=1}^N \frac{2\ell+1}{4\pi} P_\ell(\hat{j} \cdot \hat{j'}) \label{eq:fourth}\\
    &= \frac{1}{N^2} \sum_{j=1}^N \sum_{j'=1}^N  P_\ell(\hat{j} \cdot \hat{j'}), \label{eq:above}
\end{align}
\end{subequations}
where $P_\ell(\hat{j} \cdot \hat{j'})$ is the Legendre polynomial of degree $\ell$ evaluated at the value $(\hat{j} \cdot \hat{j'})$. 
To proceed from  Eq.~\eqref{eq:third} to Eq.~\eqref{eq:fourth}, we used the addition theorem for spherical harmonics.
Equation~\eqref{eq:above} has a maximum value of 1, because all Legendre polynomials have a maximum value of 1.

\section{$Q_\ell$ values and inversion symmetry} \label{section:Q3}
Many 3D environments have an inversion center, meaning that $f_i(\theta, \phi) = f_i(\pi - \theta, \phi + \pi)$ for any value of $(\theta, \phi)$.
For these environments, $Q_{\ell,i} = 0$ for all odd $\ell$ values because all odd-$\ell$ $q_{\ell m,i}$ coefficients must be zero for any spherical distribution with inversion symmetry:
\begin{align}
    f_i(\theta, \phi) &= f_i(\pi - \theta, \phi + \pi)\\
    \sum_{\ell=0}^{\infty} \sum_{m=-\ell}^\ell q_{\ell m,i} Y_{l\ell m}^*(\theta, \phi) &= \sum_{\ell=0}^{\infty} \sum_{m=-\ell}^\ell q_{\ell m,i} Y_{\ell m}^*(\pi-\theta, \phi+\pi) \label{above2}\\
    \sum_{\ell=0}^{\infty} \sum_{m=-\ell}^\ell q_{\ell m,i} Y_{\ell m}^*(\theta, \phi) &= \sum_{\ell=0}^{\infty} \sum_{m=-\ell}^\ell (-1)^\ell q_{\ell m,i} Y_{\ell m}^*(\theta, \phi) \label{above3}
\end{align}
For Eq.~\eqref{above3} to be true, $q_{\ell m,i}$ must be zero for all odd $\ell$ and associated $m$ values.
To proceed from Eq.~\eqref{above2} to Eq.~\eqref{above3}, we have used the known parity symmetry of the spherical harmonics. 

\section{The Kabsch algorithm} \label{section:kabsch}
We outline the Kabsch algorithm following     Ref.~\onlinecite{Kavraki}. 
Let $\{\bm{x}_n\}$ and $\{\bm{y}_n\}$, $n=1 \dots N$, be two sets of vectors centered at the origin. 
Let $U$ be a rotation matrix that acts on $\{\bm{x}_n\}$. The mean-squared displacement between these vector sets is
\begin{align}
E &= \frac{1}{N} \sum_n \vert U\bm{x}_n - \bm{y}_n \vert^2 . \label{eq:rewrite}
\end{align}
We can rewrite Eq.~\eqref{eq:rewrite} as a matrix equation, where $X$ and $Y$ are $3 \times N$ matrices of all vectors in $\{\bm{x}_n\}$ and $\{\bm{y}_n\}$ respectively:
\begin{subequations}
\begin{align}
NE &= \sum_n \sum_{k=1}^3 \left( UX -Y \right)_{kn} \left( UX -Y \right)_{kn}  \\
&= \sum_n \sum_{k=1}^3 \left( UX -Y \right)^T_{nk} \left( UX -Y \right)_{kn}  \\
&= \Tr \left[ \left( UX -Y \right)^T \left( UX -Y \right) \right] \label{eq:3} \\
&= \Tr X^TU^TUX + \Tr Y^TY -2\Tr Y^TUX  \label{eq:4}\\
&= \Tr X^TX + \Tr Y^TY -2\Tr Y^TUX \label{eq:5}
\end{align}
\end{subequations}
Equation~\eqref{eq:4} follows from Eq.~\eqref{eq:3} by noting that the trace of a matrix equals the trace of its transpose.
Equation~\eqref{eq:5} follows from Eq.~\eqref{eq:4}  because $U$ is an orthogonal matrix, and thus $U^T = U^{-1}$.
Minimizing $E$ means choosing $U$ such that $\Tr Y^TUX$ is maximal. 
This quantity can be thought of as the overlap between the rotated set of vectors $UX$ and the unrotated set $Y$. 
If bra-ket notation is easier to intuit, each element of the trace is equivalent to $\bra{\bm{y}_n}U\ket{\bm{x}_n}$. 
We find $U$ by performing a singular value decomposition  $XY^T = VSW^T$, where $V$ and $W^T$  are orthonormal matrices of the left and right eigenvectors of $XY^T$, and $S$ is a diagonal matrix of its eigenvalues in decreasing order:
\begin{subequations}
\begin{align}
\Tr Y^TUX &= \Tr XY^TU  \\
&= \Tr VSW^TU  \\
&= \Tr SW^TUV
\end{align}
\end{subequations}
Because $S$ is a diagonal matrix, the trace is a (weighted) sum over the diagonal elements of $W^TUV$. $W^TUV$ is an orthonormal matrix because it is a product of orthonormal matrices, and elements of $S$ are never negative, so the trace is maximal when $W^TUV=I$, the identity matrix. 
$I$ is the orthonormal matrix with maximal trace. 
Thus, the $U$ that minimizes $E$ is given by
\begin{subequations}
\begin{align}
W^TUV &= I  \label{first} \\
U &= WV^T \, . \label{eq:last2}
\end{align}
\end{subequations}
Equation~\eqref{eq:last2} follows from Eq.~\eqref{first}  because $W$ and $V$ are orthonormal: $WW^T=I$ and $VV^T=I$.
If $U$ found in this manner is an improper rotation, meaning that $\det U = -1$, we must instead use the next best (proper) rotation by setting the final column of $W^TUV$ to be $(0,0,-1)$ rather than $(0,0,1)$. 
This will insure that the next best $U$ is a proper rotation, and subtracts the smallest element of $S$ during the trace, rather than adding it. 
The optimal proper rotation $U$ can be concisely written as
\begin{align}
U &= W \begin{pmatrix} 1&0&0 \\ 0&1&0 \\ 0&0&d\end{pmatrix} V^T,
\end{align}
where $d=\text{sign}\left( \det XY^T\right)$, because $\det XY^T = \det V \det S \det W$ has the same sign as $\det U = \det W \det V$.

\bibliographystyle{apsrev4-2}
\bibliography{ajp}

\end{document}